\documentclass{ar-1col}
\usepackage{graphicx,natbib}
\usepackage{txfonts}
\usepackage[utf8]{inputenc}
\usepackage{xcolor}

\jname{Xxxx. Xxx. Xxx. Xxx.}
\jvol{54}
\jyear{2016}
\doi{10.1146/((please add article doi))}

\newcommand{\eg} {{\em e.g.}}

\newcommand{\re}{\textcolor[rgb]{1,0,0}}
\newcommand{\rev}{}
\newcommand{\revv}{}

\def\ltsima{$\;\buildrel < \over \sim \;$}
\def\simlt{\lower.5ex \hbox{\ltsima}}

\def\gtsima{$\;\buildrel > \over \sim \;$}
\def\simgt{\lower.5ex \hbox{\gtsima}}

\begin{document}

\markboth{Maryvonne Gerin, David A. Neufeld and Javier R. Goicoechea }{Interstellar Hydrides}

\title{Interstellar Hydrides}


\author{Maryvonne  Gerin,$^{1,2}$ David  A. Neufeld,$^{3,4}$\\and Javier R. Goicoechea$^5$
\affil{$^1$LERMA, Observatoire de Paris, PSL Research University, CNRS, UMR8112,Paris, France, F-75014, Paris,
 email:maryvonne.gerin@ens.fr}
\affil{$^2$Sorbonne Universit\'es, UPMC Univ. Paris 06, UMR8112, LERMA, Paris, France, F-75005 }
\affil{$^3$Department of Physics and Astronomy, The Johns Hopkins University, Baltimore, USA, MD 21218, email : neufeld@jhu.edu} 
\affil{$^4$Visiting Professor, Astronomy Department, University of Maryland, 
College Park, USA, MD20742}
\affil{$^5$Grupo de Astrof\'{\i}sica Molecular, 
Instituto de Ciencia de Materiales de Madrid (CSIC).  E-28049. Madrid, Spain. email: jr.goicoechea@icmm.csic.es}
}
\begin{abstract}
Interstellar hydrides -- that is, molecules containing a single heavy
 element atom with one or more hydrogen atoms -- were among the first
 molecules detected outside the solar system.  They lie at the root of
 interstellar chemistry, being among the first species to form in
 initially-atomic gas, along with molecular hydrogen and its associated
 ions. Because the chemical pathways leading to the formation of
 interstellar hydrides are relatively simple, the analysis of the
observed  abundances is relatively straightforward and provides
 key information about the environments where hydrides are found.  Recent
 years have seen rapid progress in our understanding of
 interstellar hydrides, thanks largely to far-IR and submillimeter observations
 performed with the {\it Herschel} Space Observatory.  
In this review, we will discuss observations of
 interstellar  hydrides, along with the advanced modeling approaches that have been used
 to interpret them, and the unique information that has thereby been
 obtained.

\end{abstract}


\maketitle

\tableofcontents

\section{INTRODUCTION} \label{sec:intro} \subsection{Historical perspective}

Although dust and gas clouds were suspected to exist early in the 20th century, the survival
of molecules in the harsh ISM was more  controversial \citep[e.g.][]{eddington:26}. Together
with CN, the hydrides CH and CH$^+$ were the first gas-phase molecules detected in the
1930's and 1940's \mbox{\citep[e.g.][]{swings:37,douglas:41}}. The improved spectral
resolution of visible light spectrometers allowed the detection of very narrow absorption
lines in the spectra obtained towards nearby and bright stars. Their associated wavelengths
were well known to molecular spectroscopists and the small line-widths suggested low gas
temperatures, much lower than those of the stellar atmospheres producing broad atomic
features. Indeed, the observed narrow lines remained stationary in the spectra of
spectroscopic binary stars whose photospheric lines showed periodic velocity shifts. Thus
they were attributed to foreground interstellar clouds (today associated with diffuse clouds
with gas densities, defined as $n_{\rm H}$=$n$(H)+2$n$(H$_2$), of less \rev{than a few hundred}
hydrogen atoms per cm$^{3}$ and gas kinetic temperatures, $T$, of about 100\,K). The
interpretation of these pioneering hydride detections was not obvious. Early attempts to
model the formation of CH and CH$^+$ focused on the slow radiative association of C$^+$ and
C with hydrogen atoms \citep{kramers:46}. Chemical reactions with H$_2$ were ignored just 
because they were known to be endothermic, and because interstellar H$_2$ was not even
detected. It was also suggested that molecules such as CH$^+$ could be a dissociation
product of CH$_4$ sublimated from grains near stars \citep{bates:51}. Both atomic and
molecular hydrogen were discovered later in the cold neutral medium (CNM), H{\sc{i}} at
1420\,MHz by radio techniques in the 1950's \citep{ewen:51},  and H$_2$ twenty years later
by electronic absorption lines in the FUV from sounding rockets and from \textit{Copernicus}
satellite observations \mbox{\citep[e.g.][]{carruthers:70,spitzer:73}}.

\begin{marginnote}[
]
\entry{ISM}{Interstellar Medium} \end{marginnote}

Millimeter- and radio-frequency searches for rotational and lambda-type doubling emissions
were also recognized as a promising method for detecting new interstellar molecules, for
example by \citet{townes:57}, who presented predictions for the frequencies of several key
transitions. These early line frequency predictions contributed to the development of radio
astronomy in the cm domain and led to the first detection of OH at 1.7\,GHz
\citep{weinreb:63}, NH$_3$ at 23\,GHz  \citep[the first polyatomic molecule in
space,][]{cheung:68} and H$_2$O at 22\,GHz \citep{cheung:69}. OH was also the first molecule
detected outside the Milky Way \citep{weliachew:71}. NH$_3$ and H$_2$O were discovered by
the team of Nobel Prize winner C.H. Townes. Most of the subsequent discoveries of new
molecules required the development of higher frequency heterodyne receivers in the mm and
submm domains available from ground-based telescopes \citep[e.g.,][]{falgarone:05}, and more
recently in the far-infrared (FIR), a wavelength region that is only available from stratospheric and space
observations \citep[e.g.,][and references therein]{gerin:12}. These techniques allow the
detection of pure rotational transitions at very high spectral resolution (less than
1\,km\,s$^{-1}$); thus fine and hyperfine line splittings can be  resolved. In addition, the
FIR domain allows the detection of rotationally excited emission lines. This permits high temperature
and density environments (\mbox{$T$$>$100\,K} and \mbox{$n_{\rm H}$$>$10$^{4}$\,cm$^{-3}$}),
such as those in star forming regions, to be probed. In diffuse and translucent clouds,
hydrides such as CH$^+$, CH, HCl, NH, OH, OH$^+$ or SH can be nowadays also studied through
their electronic absorption bands in the UV-visible domain  at velocity resolutions of a few
km\,s$^{-1}$ \citep[e.g.,][]{federman:95,krelowski:10,zhao:15}. NH in particular, was first
detected in this domain \citep{meyer:91}. In dense but cold clouds
(\mbox{$T$$\simeq$20\,K}), hydrides also exist in the solid-phase as ice mantles covering
dust grains \rev{\citep[see the review by][]{boogert:15}}. 
Their vibrational modes can be detected in the IR domain. The first mantles
detected in absorption were water and ammonia ices \citep[][]{merrill:76}. \mbox{Table~1}
shows a compilation of the detected hydrides and the wavelength range in which they  have
been observed.

\begin{marginnote}[
]
\entry{FUV}{Far ultraviolet, defined as the 91.2 to 200~nm wavelength region, corresponding
to  photon energies between 6 and 13.6~eV} \end{marginnote}

\begin{marginnote}[
]
\entry{FIR}{Far infrared, defined as the 20 to 300$\mu$m wavelength region, corresponding to
frequencies between 1 and 15 THz.} \end{marginnote}

The formation of H$_2$ on grain surfaces rather than in the gas-phase was soon established,
and first models suggested that most hydrogen should be molecular at visual extinctions
$A_V$$>$1.5\,mag \mbox{\citep[e.g.,][]{hollenbach:71}}. Other molecules were proposed to
form on grains as well \citep{watson:72}. More complete chemical models of diffuse clouds
were developed \citep{solomon:72, black:77}. The chemistry of molecular cloud interiors
(most hydrogen in H$_2$) was also modeled and predicted to be dominated by ion-molecule
reactions maintained by cosmic-ray particle ionization \citep{herbst:73}. These chemical
models were the basis of more sophisticated diffuse cloud, dense photodissociation region
(PDR), shock and turbulence dissipation models developed later in the 1980's. Supported by
laboratory and theoretical studies, these models were also critical in predicting  the
existence of new hydride molecules in space. Review articles have been presented previously by 
\citet{ewine:98} and \citet{snow:06}  on diffuse interstellar clouds, by  \citet{hollenbach:99} on PDRs, and 
by \citet{bergin:07} and  \citet{caselli:12b} for dark clouds. 

Recent years have seen rapid progress in our understanding of interstellar hydrides, with
new detections  and modeling approaches demonstrating their diagnostic power. Thus it seems  
appropriate to review the current state of observations  in different interstellar
environments, and the unique astrophysical information that can be obtained. This review is
organized as follows. After a general introduction of the thermochemical properties and 
of the excitation of hydrides, in Section~2 we summarize the main gas-phase
and solid-phase processes driving the chemistry of hydrides. 
We then focus on the appearance of hydride
molecules in the diffuse molecular gas (Section~3), in dense and strongly UV-irradiated gas (Section~4),
in shocks and turbulent dissipation regions (Section~5), and 
in cold and dense gas shielded from UV radiation (Section~6).  
In Section~7 we summarize the diagnostic power of hydrides, and in Section~8
we extend our discussion to external galaxies.  We end by giving our conclusions 
and anticipating some of the  exciting prospects for hydride research for the years to come.

\begin{marginnote}[
]
\entry{PDR}{Photo-Dissociation Region} \end{marginnote}

\begin{table}
\tabcolsep7.5pt
\caption{Main astrophysical hydrides$^{\rm a)}$}
\label{tab:list}
\begin{center}
\begin{tabular}{@{}l|c|c|c@{}}
\hline
Formula & Name & Spectral domain$^{\rm b)}$  & Ref.\\
\hline
H$_2$ & Molecular Hydrogen & UV-Visible, IR, FIR &  \cite{carruthers} \\
H$_3^+$ & Protonated molecular hydrogen & IR & \cite{geballe:96}\\
\hline
CH & Methylidyne & UV-Visible, (sub)mm, cm & \cite{swings:37}\\
CH$_2$ & Methylene & FIR, (sub)mm &\cite{hollis:95}\\
CH$_3$ &Methyl  & IR &\cite{feuchtgruber:00}\\
CH$_4$ & Methane & IR & \cite{lacy:91}\\
CH$^+$ & Methylidynium & UV-Visible, FIR, (sub)mm & \cite{douglas:41}\\
\rev{CH$_3^+$} & \rev{Methylium} & \rev{ IR, (sub)mm} & \cite{roueff:13}$^{\rm c}$\\
\hline
NH & Imidogen & UV-Visible, (sub)mm & \cite{meyer:91}\\
NH$_2$ & Amidogen & (sub)mm &  \cite{vandishoeck:93}\\
NH$_3$ & Ammonia & (sub)mm, cm& \cite{cheung:68}\\
NH$_4^+$ & Ammonium & (sub)mm & \cite{cernicharo:13}$^{\rm d)}$\\
\hline
OH & Hydroxyl radical & UV-Visible, FIR, cm  & \cite{weinreb:63}\\
H$_2$O & Water       & FIR, (sub)mm, cm    &  \cite{cheung:69}\\
OH$^+$ & Hydroxylium & UV-Visible, (sub)mm   & \cite{wyrowski:10}\\
H$_2$O$^+$ & Oxidaniumyl &UV-Visible, (sub)mm &\cite{ossenkopf:10}\\
H$_3$O$^+$& Hydronium & FIR, (sub)mm & \cite{phillips:92}\\
\hline
HF & Hydrogen fluoride & FIR & \cite{neufeld:97}\\
\hline
SH & Mercapto radical & UV-visible, FIR & \cite{neufeld:12}\\
H$_2$S &Hydrogen sulfide  & (sub)mm & \cite{thaddeus:72} \\
SH$^+$ & Sulfanylium & (sub)mm & \cite{benz:10,menten:11}\\
\hline
HCl & Hydrogen chloride & UV-visible, (sub)mm & \cite{blake:85}\\
HCl$^+$ & Chloroniumyl  & FIR & \cite{deluca:12}\\
H$_2$Cl$^+$ & Chloronium & (sub)mm & \cite{lis:10}\\
\hline
ArH$^+$ & Argonium & (sub)mm & \cite{barlow:13}\\
\hline
\end{tabular}
\end{center}
\begin{tabnote}
$^{\rm a)}$Adapted from The Astrochymist (www.astrochymist.org); $^{\rm b)}$ The corresponding wavelength ranges are : UV-Visible 100 -- 1000nm, IR : 1 -- 20 $\mu$m, FIR 20 -- 300 $\mu$m ; (sub)mm 0.3 -- 4 mm; cm 1 - 20 cm; $^{\rm c)}$ A tentative detection of the isotopologue CH$_2$D$^+$ is reported;  $^{\rm d)}$ The detection of the 
isotopologue NH$_3$D$^+$ is reported.
\end{tabnote}
\end{table}

\subsection{Thermochemistry of interstellar hydrides}
\label{sec:thermo}
Although the interstellar gas is far from thermochemical equilibrium, key features of the chemistry of interstellar hydrides -- discussed in detail in Section \ref{sec:chemistry} below -- are determined by thermochemistry 
\citep{neufeld:09}.  Among the elements in the second and third rows of the periodic table, F, O, N, Ne and Ar all have first ionization potentials greater than that of atomic hydrogen (13.60 eV); because hydrogen atoms shield them very effectively from ultraviolet radiation of sufficient energy to ionize them, these elements are predominantly neutral in the CNM.   All other elements in the second and third rows of the periodic table -- including C, Si, P, S and Cl -- have ionization potentials smaller than that of hydrogen; typically,
they are predominantly singly-ionized in the cold diffuse ISM.  (The exception is Cl, which may exhibit a significant neutral fraction in clouds with a sufficient H$_2$ fraction: this behavior is a consequence of the rapid reaction of Cl$^+$ with H$_2$ to form HCl$^+$, which subsequently undergoes dissociative recombination to form atomic chlorine).

\begin{marginnote}[
]
\entry{CNM}{Cold Neutral Medium. Two neutral phases coexist at pressure equilibrium in the ISM, the warm neutral medium with $T \sim 10^4$~K and $n_{\rm H} \sim 0.3 - 1\,$cm$^{-3}$ and the CNM with $T \sim 10^2$~K and $n_{\rm H} \sim 30 - 100\,$cm$^{-3}$.}
\end{marginnote}

\begin{figure}
\includegraphics[width=5in]{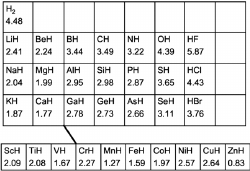}
\caption{Dissociation energies of the neutral diatomic hydrides, in eV, from \cite{neufeld:09} and based upon the thermochemical data referenced therein.}
\label{fig:NW09}
\end{figure}

With the notable exception of hydrogen fluoride, every neutral diatomic hydride has a dissociation energy smaller than that of H$_2$, as indicated in \textbf{Figure \ref{fig:NW09}}.  As a result, hydrogen abstraction reactions of the type $\rm X + H_2 \rightarrow XH + H$ are all endothermic except in the case where X is fluorine. HF is therefore the only neutral diatomic hydride that can be formed at low temperature through the reaction of an atom with H$_2$.  As for the diatomic hydride {\it cations}, HeH$^+$, OH$^+$, HF$^+$, HCl$^+$, ArH$^+$ and NeH$^+$ have dissociation energies larger than that of H$_2$.  They can therefore be produced exothermically\footnote{For Ne$^+$ and He$^+$, the alternative reaction channel to $\rm X + H + H^+$ is also exothermic and is favored.} via the hydrogen abstraction reaction $\rm X^+ + H_2 \rightarrow XH^+ + H$; such reactions are endothermic for all other ions $\rm XH^+$.   

In addition to the ionization potential of each atom, X, and the dissociation energies of the hydrides, XH and XH$^+$, another important thermochemical parameter is the proton affinity of X.  Because $\rm H_3^+$ is produced efficiently following cosmic-ray ionization of H$_2$, 
proton transfer from $\rm H_3^+$ can be an important production mechanism for $\rm XH^+$ when the proton affinity of X exceeds that of H$_2$ (422.3 kJ mol$^{-1}$ or equivalently 4.38 eV).  This condition is satisfied for C, O, Si, P, S and Cl -- but not for N, F, Ne or Ar -- with the result that \rev{the} first six atoms can react exothermically with $\rm H_3^+$ to form XH$^+$.  This consideration is most important for O, because it is the only atom among these six to be predominantly neutral in diffuse clouds.

\begin{figure}
\includegraphics[width=5in, angle=0]{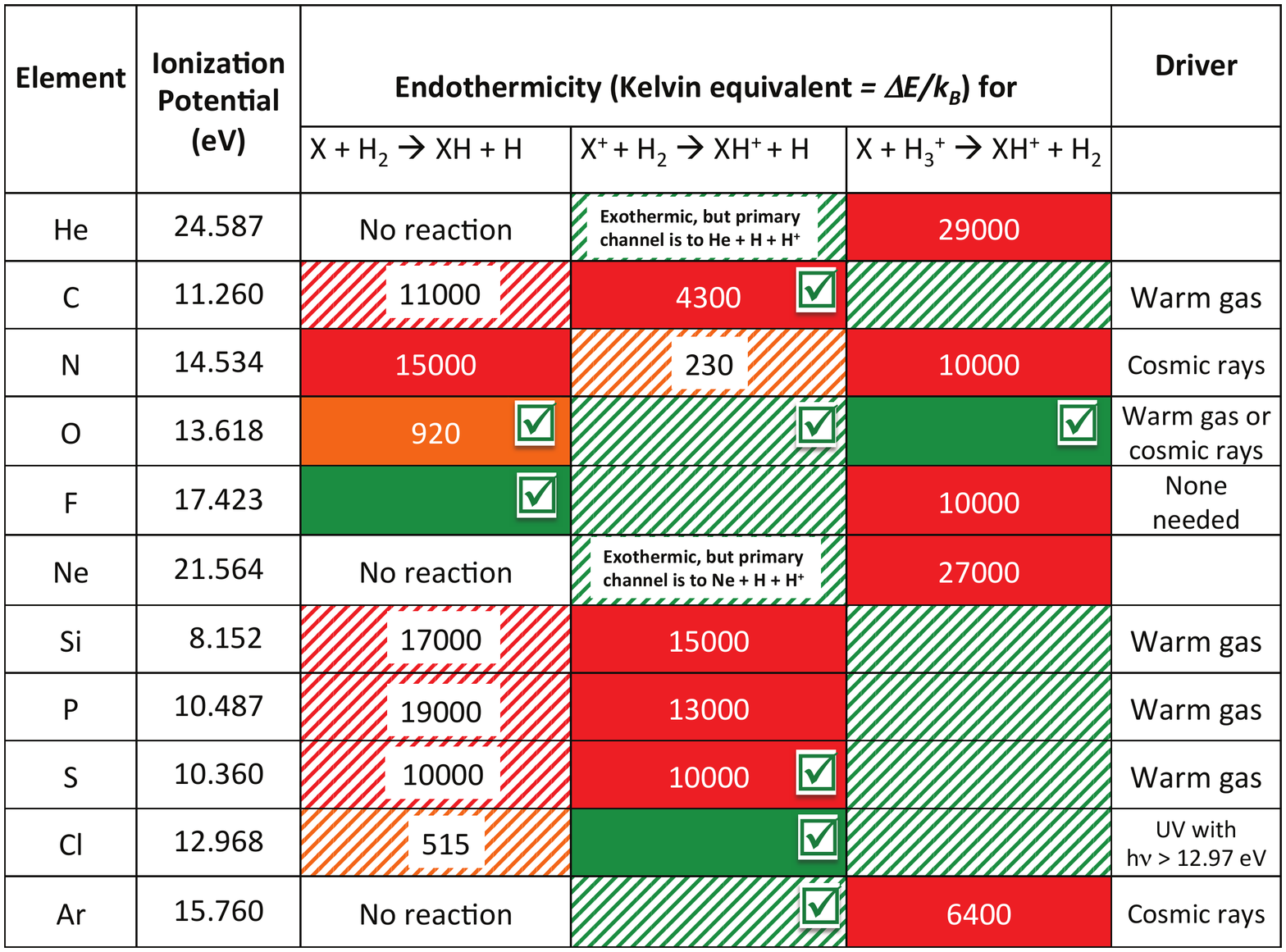}
\caption{Ionization potentials for selected elements, and endothermicities of 
three key reactions of possible importance in the formation of interstellar hydrides. See text for a complete description. \rev{As indicated in Table~\ref{tab:list}, CH, CH$^+$, NH, OH, OH$^+$, HF, SH, SH$^+$, HCl, HCl$^+$ and ArH$^+$ have been detected in the ISM.} }
\label{fig:thermochem}
\end{figure}

In \textbf{Figure \ref{fig:thermochem}}, the thermochemical considerations discussed above are presented in graphical form.  For each element, the first ionization potential is given in the second column from the left.  The 3rd, 4th, and 5th columns indicate the endothermicity of three key reactions: the reaction of X with H$_2$ to form XH, the reaction of X$^+$ with H$_2$ to form XH$^+$, and the reaction of X with H$_3^+$ to form XH$^+$.  Here, the endothermicities, $\Delta E$, are presented as temperatures, $\Delta E/k_{\rm B}$, and given in units of Kelvin.  Cells shown in green indicate exothermic reactions, those shown in orange indicate slightly endothermic reactions with $\Delta E/k_{\rm B} \le 1000$~K, and those shown in red indicate endothermic reactions with $\Delta E/k_{\rm B} \ge 1000$~K.  Particularly for neutral-neutral reactions, the activation energy (i.e.\ energy barrier) may exceed the endothermicity.  For example, \rev{the} reaction of O with H$_2$ to form OH and H has an endothermicity, $\Delta E/k_B \sim 920\rm \, K$ but an activation energy of $\Delta E_A/k_B \sim 3000\rm \, K$.
Where the reactant (X or X$^+$) is the dominant ionization state of the element in diffuse interstellar gas clouds with a small molecular fraction, 
the green, orange or red coloring is solid; otherwise, the cell is shaded.  
The dominant production mechanisms for diatomic interstellar hydrides, and the environments in which they are found, can largely be inferred from the information presented in this matrix.  Check marks indicate reactions that have been invoked as significant production routes in astrochemical models in the literature.  In some cases, different mechanisms are important in different environments: in the case of oxygen for example, production of OH via the endothermic reaction of O and H$_2$ becomes rapid in warm ($T \ge 400$~K) gas \citep[e.g.][]{kaufman:96}, while OH$^+$ production is effective in cold diffuse gas through reaction of O$^+$ with H$_2$ or (in slightly denser gas with a larger H$_2$ fraction) of O with H$_3^+$ \citep[e.g.][]{hollenbach:12}.

\subsection{Excitation of interstellar hydrides}
\label{sec:excitation}
Several physical processes determine the excitation of interstellar hydrides \citep[e.g.][and references therein]{Roueff:13b}, including collisional excitation, spontaneous radiative decay, radiative excitation, and formation pumping.   At sufficiently high densities, the first of these processes drives the molecular excitation to local thermodynamic equilibrium (LTE),  in which the relative level populations are given by Boltzmann factors determined by the gas kinetic temperature.  However, the ``critical density" needed to achieve LTE is typically larger for the rotational states of hydrides than that for those of non-hydride interstellar molecules, because hydrides have smaller momenta of inertia, larger rotational constants, and much larger spontaneous radiative rates (the latter scaling as the cube of the transition frequency for a given dipole matrix element).  At densities below the critical density, the level populations reach a quasi-equilibrium in which the total production and loss rates are equal for every state.  In general, the excitation is ``sub-thermal," with the excitation temperatures for most transitions being smaller than the gas kinetic temperature.  However, specific transitions can sometimes be supra-thermally-excited, or even inverted, giving rise to maser amplification.

\begin{marginnote}[
]
\entry{LTE}{Local Thermodynamic Equilibrium, where the level populations of a given molecule are determined by Boltzmann equilibrium at the local kinetic temperature, $T$.}
\end{marginnote}

\begin{marginnote}[
]
\entry{critical density}{The density at which the rate of collisional de-excitation from a given state is equal to the rate of spontaneous radiative decay.}
\end{marginnote}

In diffuse molecular clouds, the molecular excitation is often determined primarily by the radiation field; in clouds that are not close to a strong submillimeter radiation source, the cosmic microwave background radiation (CMB) dominates and most rotational transitions have an excitation temperature close to that of the CMB.  For molecules such as HF and HCl in which there are no metastable rotational states, this means that most molecules are found in the ground rotational state.  For triatomic hydrides such as $\rm H_2O$, $\rm H_2O^+$, and $\rm H_2Cl^+$ in which there are two spin symmetries (ortho with total H nuclear spin 1, and para with total H nuclear spin 0), most molecules in diffuse clouds are found in {\it one or other} of the lowest rotational states of the two spin symmetries.   For symmetric top molecules such as $\rm NH_3$ and $\rm H_3O^+$, in which the rotation is characterized by the quantum numbers $J$ and $K$, the lowest energy state for any given $K$ is metastable.  Thus for $\rm NH_3$ and $\rm H_3O^+$ in diffuse clouds, the population is largely divided amongst those metastable states lying at the bottom of each $K$-ladder.  

While the CMB radiation is ubiquitous, other radiation fields can become important in specific environments: strong mid-IR \rev{and FIR} radiation is present close to massive protostars, for example, and strong optical and UV radiation is present close to hot stars.  Such radiation may affect the rotational populations in the ground vibrational state, as a result of pumping \rev{through rotational (FIR),} vibrational (mid-IR) and electronic (optical and UV) transitions \citep{godard:13}.
 
In different astrophysical environments, inelastic collisions with H$_2$, with H and/or with electrons can be important or even 
dominant in determining the equilibrium populations in the various rovibrational states of an interstellar hydride.  Collisions with He can also be important but probably 
never play a dominant role.
In dense, well-shielded molecular clouds, H$_2$ is the most important collision partner, but in photodissociation regions where the carbon is significantly photoionized, excitation by electrons can be important or dominant.  While the electron abundance in such regions is only $\sim \rm few \times 10^{-4}$ relative to H$_2$, the rate coefficients for electron impact excitation can exceed those for excitation by H$_2$ or H by four orders of magnitude, particularly in the case of hydride cations or neutral hydrides with large dipole moments: here, the cross-sections are enhanced by long-range Coulomb interactions (and the mean electron velocities exceed the mean H$_2$ velocities by the square root of the H$_2$ to electron mass ratio).  In certain environments, and in particular the molecule reformation region behind dissociative shock fronts (see Section \ref{sec:shocks} below), excitation by atomic hydrogen can also be very important.

There is a considerable and ongoing effort to calculate state-to-state rate coefficients for various hydrides of astrophysical interest, most importantly in recent years for the excitation of water by ortho- and para-H$_2$ \citep[and references therein]{daniel:11} and by atomic hydrogen \citep{daniel:15}.  State-of-the-art quantal calculations make use of the so-called close-coupling method and rely upon accurate determinations of the potential energy surface for the system.  Such calculations are computationally-expensive, and are often (e.g.\ in the case of water) limited to a subset of the rovibrational states that have actually been observed in interstellar hydrides.   As a result, less computationally-intensive (and less accurate) methods such as quasi-classical trajectory calculations (e.g.\ Faure et al.\ 2007) must sometimes be used to interpret astronomical observations, in some cases accompanied by extrapolation methods involving propensity rules \citep[e.g.][]{faure:07} or the use of artificial neural networks \citep{neufeld:10b}.  The LAMDA \citep{schoier:05} and BASECOL \citep{dubernet:13} databases have been established to collate collisional rate coefficients and spontaneous radiative decay rates for hydrides and other interstellar molecules in a standard format for use in modeling the excitation.

Formation pumping is a further process that can affect the excitation state of interstellar hydrides.  Here, if the lifetime of the molecule is comparable to -- or smaller than -- the timescale on which the quasi-equilibrium level populations are established, the observed populations may reflect the initial conditions at formation.  The most clearly-established case of formation pumping in interstellar hydrides is the excitation of high-lying OH rotational states following the photodissociation of H$_2$O to form OH \citep[discussed further in Section \ref{sec:shocks} below]{tappe:08}.  Formation pumping has also been discussed as an explanation for the observed excitation of H$_3$O$^+$ \citep[e.g.][]{lis:14,gonzalez:13}, and may be important for other molecular ions as described further in
 Section \ref{sec:pdr} below.

\section{CHEMISTRY}
\label{sec:chemistry}
\subsection{Gas phase processes}

\begin{figure*}
\centering
\rotatebox{0}{\resizebox{!}{8cm}{
\includegraphics{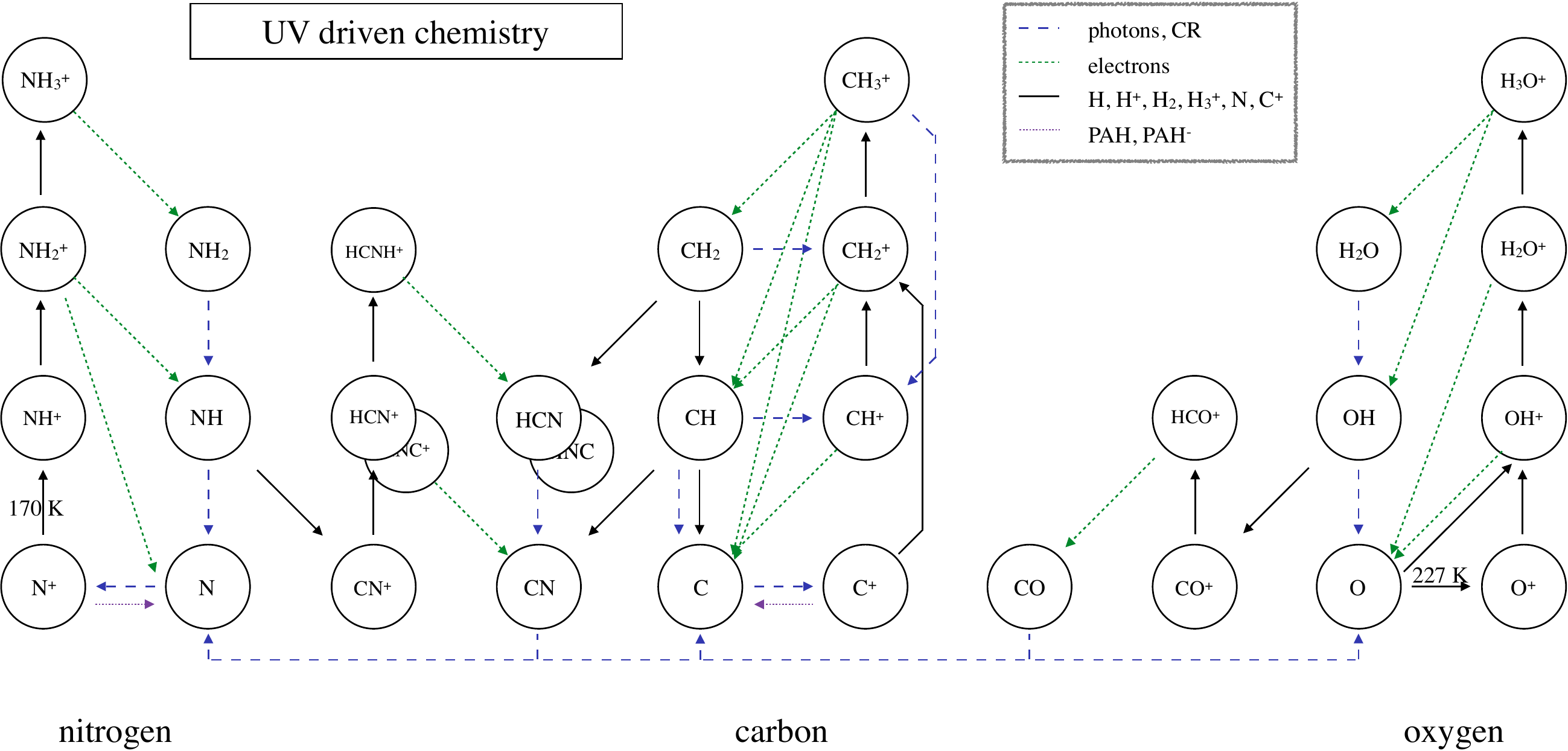}
}}
\caption{\rev{Illustration of the chemical network initiating the carbon, oxygen and nitrogen chemistry in diffuse cloud conditions ($n_{\rm H} =  50$~cm$^{3}$, $A_V = 0.4$~mag, $\chi = 1$). The black arrows show the reactions with H, H$^+$, H$_2$, H$_3^+$, C$^+$ and N, with values of the endothermicity for the reaction between N$^+$ and H$_2$ and for the charge exchange reaction between O and H$^+$. Note that CH$_2^+$ is formed in the slow radiative association  reaction between C$^+$ and H$_2$. The dashed blue arrows indicate the reactions induced by far ultra violet (FUV) photons or cosmic rays. Dissociative recombination reactions with electrons are shown with green dotted arrows. Purple arrows show the neutralization reactions on dust grains and PAHs. Adapted from \cite{godard:14}. \label{fig:network-co}}}
\end{figure*}

As anticipated in Section \ref{sec:thermo}, hydrides are mostly formed 
 in the gas through series of hydrogen abstraction reactions
combined with dissociative recombination. The chemistry of carbon,  
oxygen and nitrogen hydrides is illustrated in 
\textbf{Figure \ref{fig:network-co}}. 
This relative simplicity hides 
 subtleties introduced by {\sl i)} the presence of highly endothermic 
reactions on important pathways (e.g. the initiation reaction of the
carbon chemistry between ionized carbon and molecular hydrogen has an endothermicity of 0.37~eV or 4300~K), and {\sl ii)}
the dependence of the reaction rates and products on the molecular 
hydrogen spin symmetry state, \rev{or more generally on the initial rovibrational state (for molecules) or the fine structure level (for atoms) of the reactants \citep[e.g.][]{faure:13,li:13}}. 
Fortunately, many of the key reactions for hydride synthesis have been 
studied through  theoretical calculations and laboratory measurements because
of their fundamental interest for molecular physics. 
For example, the neutral-neutral reaction between F and H$_2$ forming HF and H is one of the benchmark systems for the tunneling effect. Combining measurements down to 11~K and new theoretical calculations, \citet{tizniti:14}  have provided an accurate determination of this important reaction rate.

The hydrogen abstraction reactions initiating the 
 oxygen, nitrogen, chlorine and argon chemistry are exothermic or moderately 
endothermic (for N). 
Because dissociative recombination reactions are fast, photodissociation is
usually not a dominant destruction process for the hydride ions, while
it affects neutral hydrides. 
The hydride chemistry is sensitive to the \rev{strength and spectrum of the} incident X-ray and UV radiation field, and
to the ionization rate due to cosmic rays, the two main sources of ionized
species. The exact energy distribution of the UV field is particularly important for the
chlorine chemistry since the ionization potential of atomic chlorine, 12.968~eV,
is very close to the Lyman edge (13.6~eV). \rev{Polycyclic Aromatic Hydrocarbons (PAHs) play a role
in the chemistry because cations (e.g. H$^+$, C$^+$, N$^+$) get neutralized when encountering a neutral or
negatively charged PAH. Hence PAHs reduce the efficiency of molecule formation. This is especially important for
the oxygen chemistry in which the initiating charge exchange reaction between H$^+$ and O competes
with neutralization on PAHs. }

\begin{marginnote}[
]
\entry{Cosmic Rays (CR)}{Cosmic Rays are energetic particles (electrons, protons, and heavier nuclei), accelerated at energies between 0.1 and 10$^6$ GeV in fast shocks.}
\end{marginnote}

\begin{marginnote}[]
\entry{PAHs}{Polycyclic Aromatic Hydrocarbons are carbon and hydrogen molecules composed of multiple benzenic cycles with hydrogen atoms at the boundaries.}
\end{marginnote}

Given the relatively large reaction rates,  the  chemical time scales
for gas phase hydride production are short, typically less than a few 100 yr 
provided molecular hydrogen has reached its equilibrium abundance. The formation
of molecular hydrogen itself, which occurs on grain surfaces, is significantly slower, with typical 
time scales of  10 Myr for diffuse gas densities. Numerical simulations
of the diffuse ISM, using either hydrodynamical or magnetohydrodynamic (MHD) codes,
 have shown however that the equilibrium molecular hydrogen 
fraction  can be reached faster in a dynamical medium
 because  the  compressions induce transient density enhancements where
H$_2$ forms more rapidly than in an uniform medium \citep{glover:07b}.
  The time scale for conversion between the H$_2$ spin symmetry states
 strongly depends whether only gas phase processes
contribute, or whether H$_2$ is rapidly thermalized on dust grains 
\citep{lebourlot:00}. In the former case, the conversion is slow, with a
similar time scale as the static formation rate of H$_2$  \citep[e.g.][]{pagani:11}. Conversion on
grain surfaces is  however expected to be faster, but a small fraction of $o$-H$_2$ \rev{($\sim 10^{-3}$)} is expected to remain even at low temperatures because of the presence of newly formed H$_2$ molecules, either on grains or through chemical reactions \citep{lebourlot:00}. Several hydrides (e.g.  NH$_3$, NH$_2$ or H$_2$O) exhibit spin symmetry states whose 
relative abundances are sensitive to the H$_2$ ortho-to-para ratio (OPR) in the gas.

\begin{marginnote}[
]
\entry{OPR}{Ortho-to-para ratio, the ratio between the densities, or column densities, of the ortho and para nuclear spin symmetry states.}
\end{marginnote}

\begin{figure}
\includegraphics[width=16cm]{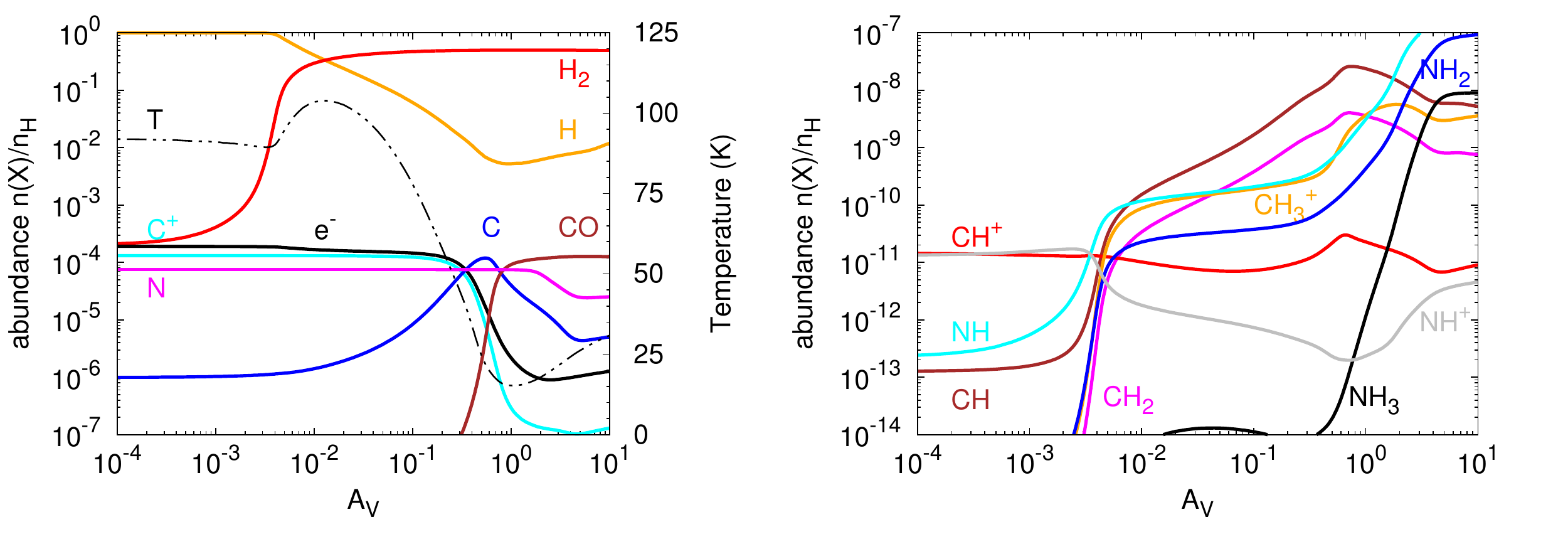}
\caption{\label{fig:prediction}Variation of the gas phase abundances of various species relative to the total hydrogen content as a function of depth $A_V$ for a cloud with constant pressure $P/k_B = 4\times 10^3$~K\,cm$^{-3}$, irradiated by the average interstellar radiation field ($\chi = 1$) \rev{from the left side}, and a cosmic ray ionization rate $\zeta_{\rm H} = 2 \times 10^{-16}$ s$^{-1}$, computed with the Meudon PDR code \citep{LeBourlot:12}. The left panel displays 
the hydrogen (H, H$_2$), carbon (C$^+$, C, CO), nitrogen (N, N$_2$ not shown) reservoirs together with the electron abundance and the kinetic temperature (right vertical scale). The gas density ranges from about 40 cm$^{-3}$ at A$_V \sim 0.01$~mag up to $\sim 150$~cm$^{-3}$ at $A_V > 1$~mag. The right panel shows the
main carbon and nitrogen hydride abundances as a function of extinction. }
\end{figure}

State of the art chemical models predict with a good accuracy (better than a factor of two) the observed abundances of \rev{most of the} abundant hydrides, and their respective
relationships. \textbf{Figure~\ref{fig:prediction}} illustrates the behavior of 
carbon and nitrogen hydrides for a constant pressure cloud with $P/k_B = 4\times10^3$~K\,cm$^{-3}$ \rev{using a pure gas-phase chemical network (except for H$_2$).}  The  behavior of oxygen hydrides is illustrated in the left panel of \textbf{Fig.} \ref{fig_pdr_mod_hol} (Section \ref{sec:pdr}) which displays a model with a similar value of the control parameter $\chi/n_{\rm H}$, where $\chi$ represents the enhancement factor of the incident FUV radiation field over the mean interstellar radiation field in the solar neighborhood.   
It is interesting to note that CH and CH$_2$ show a very similar behavior with
a first rise of their relative abundances   nearly coinciding with the H/H$_2$ transition at A$_{\rm V} \sim 0.1$ for this particular model. The appearance of nitrogen hydrides (NH, NH$_2$ and NH$_3$) 
takes place further inside the cloud at A$_{\rm V} \sim 1$, and coincides with the CO formation zone. The
NH$^+$ abundance remains low, in agreement with {\sl Herschel} observations \citep{persson:12}. 
The series of molecular ions OH$^+$, H$_2$O$^+$ and H$_3$O$^+$ behaves somewhat differently from the carbon and nitrogen 
hydrides : each ion is predicted to probe a different zone of  molecular hydrogen content, with  $f_{\rm H_2} = 2n({\rm H}_2)/[2n({\rm H}_2)+n({\rm H})]$, the molecular fraction, increasing from OH$^+$ to H$_3$O$^+$. As  
will be  discussed further in Section \ref{sec:diagnostic} below, this property is fully consistent with observations and can be used to derive the fraction of molecular hydrogen in the region of maximum OH$^+$ abundance.

\subsection{Solid phase processes}

Molecular hydrogen is not the only molecule formed in the solid phase. Recent 
studies have shown, that even under diffuse ISM conditions, 
the formation of molecules
on grain surfaces can be efficient \rev{ \citep[see e.g.][ for H$_2$O]{sonnentrucker:15}} .  
Hydrides are particularly susceptible to
being formed on grain surfaces because of the ubiquitous presence of hydrogen.
Laboratory experiments using  \rev{beams} of hydrogen and oxygen atoms
 hitting a cold surface have clearly demonstrated that water ice forms in this process. A  fraction of the newly formed molecules does not
stay on the cold surface but is immediately released in the gas phase 
\citep{dulieu:13}, a process called chemical desorption. The fractions of trapped versus ejected molecules depend
 on the nature of the surface, the ice coverage, and the exact reaction mechanism. 
In a similar manner as water ice, the hydrogenation of carbon and nitrogen atoms
produces methane and ammonia. Laboratory experiments are now able to
 probe these processes  including exchanges between H and D atoms and the production of deuterated species \citep{fedoseev:15}.

Once formed on the cold grain surfaces, four mechanisms may release 
molecules to the gas phase: {\sl i)}
\rev{thermal sublimation when the grains reach temperatures larger than the ice sublimation threshold},  {\sl ii)} photo-desorption when FUV photons are absorbed by the grain surface and kick some
molecules out of the \rev{ice}, {\sl iii)} cosmic ray induced desorption when an energetic particle
interacts with a \rev{grain mantle}, and {\sl iv)} chemical desorption during the formation process. Laboratory experiments have been performed to
study these mechanisms and understand in which conditions they operate. 
 \rev{Theoretical and  laboratory studies of ice  photodesorption have concluded that this
  process is complex. FUV photons can dissociate a frozen water molecule producing two fragments (H and OH) that have excess energy.  These fragments can either recombine as a gas phase water molecule, or the excited H can kick out another frozen water molecule. The outcome  depends on many parameters, first of all where the  FUV photon is absorbed, since only the upper ice layers participate to the desorption. The presence of other isotopologues, of other molecules like CO, which does not dissociate upon the absorption of a FUV photon, will also affect the desorption yield.}
Overall the 
photodesorption yields range between 10$^{-4}$ and $10^{-2}$ molecules/photon \citep{arasa:15}. With such numbers, most of the species
are predicted to be frozen in dense core conditions, with a minor
fraction remaining in the gas phase. As illustrated in Fig. \ref{fig_pdr_mod_hol}, water vapor reaches its peak abundance for extinctions of a few magnitudes.  
 Chemical models also include a temporal dimension because  the dynamical time scales, which scale with the free fall time $t_{ff} = \sqrt{3\pi / 32G\rho} \sim 10^5 \sqrt{5\times 10^4{\rm cm}^{-3}/n({\rm H_2})}$~yr, are similar to the chemical time scales. For instance, the \rev{freeze-out time scale in the absence of desorption is
 $t_{freeze} \sim 10^5 \big( 10^4 \rm{cm}^{-3}/n({\rm H_2}) \big) $~yr.}

While the respective abundances of H$_2$O, CH$_4$ and NH$_3$ are fairly well documented in interstellar ices, little is known about the presence
of hydrides composed of the heavier elements F, S or Cl in ices. The somewhat lower elemental abundances  precludes a direct identification given the limited sensitivity of infrared spectroscopy of solid features, 1\% of the water ice content. 
Solid phase and gas phase processes are coupled, as shown for instance by the production of H$_2$ and  methane during the FUV irradiation of hydrogenated amorphous carbons \citep{alata:14}. 
Although laboratory experiments have led to steady progress in the understanding of the solid phase chemistry, the state of the art is still far from that of the chemistry in gas phase, even for simple hydrides. \rev{The 
relative contributions of solid phase and gas phase routes to the formation of nitrogen hydrides deserves specific attention. While the reaction between N$^+$ and H$_2$ initiates the nitrogen chemistry 
in  UV-illuminated regions, NH is mostly formed as 
as a minor channel in the dissociative recombination of  N$_2$H$^+$ in shielded regions. As discussed by  \cite{wagenblast:93}, the gas-phase production route for nitrogen hydrides is not efficient enough to explain the observed abundance in diffuse clouds, especially if the cosmic ray ionization rate is low. An alternative  formation mechanism for  nitrogen hydrides involves the hydrogenation of nitrogen atoms  on grain surfaces to form ammonia, and its release into the gas phase for further processing. Given the wealth of data now available,  the
relative contributions of solid phase and gas phase routes in the nitrogen chemistry should be  assessed, especially for diffuse conditions.  More generally, more} quantitative evaluations of the reactivity of atoms and molecules  (including  diffusion)  and of the efficiency of the various desorption processes
is needed to improve astrophysical models.

\section{HYDRIDES IN DIFFUSE GAS}
\label{sec:diffuse}
\subsection{Overview}

With the exception of  CH$_4$, CH$_3^+$ and NH$_4^+$, all the hydride molecules listed in Table \ref{tab:list} are known to be present in diffuse molecular clouds.  Observations of hydrides in diffuse clouds have targeted transitions over a wide range of wavelengths, from a 122~nm electronic transition (of OH) to a set of 18~cm lambda-doubling transitions (also of OH). Typically, molecules in diffuse clouds have been observed in absorption toward background sources of continuum radiation; these include hot stars (at ultraviolet wavelengths), warm dust in regions of active star formation (at FIR and submillimeter wavelengths), and HII regions (at centimeter wavelengths).  With the exception of those targeting centimeter-wavelength lambda-doubling transitions, absorption-line observations yield relatively robust and model-independent estimates of molecular column densities.  This advantage results from the fact -- discussed previously in Section \ref{sec:excitation} -- that under typical conditions in diffuse clouds, most hydride molecules are found primarily in the ground state (or, for triatomic hydrides, in the lowest energy states of the ortho- and para-symmetry states); thus, stimulated emission is typically negligible, and the scaling between absorption line optical depth and molecular column density is only weakly dependent upon the physical conditions.  In the remainder of this section, the observational data on hydride molecules in diffuse clouds will be reviewed.
In Section \ref{sec:diagnostic} below, we will discuss how the measured abundances of interstellar hydrides can be used as quantitative probes of the environment in which they are found.

\subsection{Optical/UV and centimeter-wavelength observations of interstellar hydrides}

In diffuse molecular clouds, several hydride molecules show observable absorption at optical and ultraviolet wavelengths through their electronic transitions.  These include three molecules -- 
CH, CH$^+$, and NH -- that were first detected (see Table \ref{tab:list}) in \rev{this} spectral region; 
two molecules -- OH$^+$ \citep{krelowski:10} and SH \citep{zhao:15} -- with optical/UV transitions that were  detected shortly after their initial discovery at submillimeter wavelengths; as well as OH, detected both at vacuum UV \citep[e.g.][]{snow:76} and near-UV wavelengths \citep[e.g.][]{porras:14}.  In this spectral region, spectral resolving powers $R = \lambda / \Delta \lambda$ in excess of $\sim 10^5$ can be achieved with echelle spectrometers.  Although not as high as those achievable at submillimeter wavelengths using heterodyne receivers, these spectral resolving powers yield valuable kinematic information about the absorbing molecules.  For example, a recent analysis of near-UV spectra showing OH$^+$, OH, CH, CH$^+$, and CN absorption \citep{porras:14} has revealed a kinematic association between OH$^+$, CH$^+$, and neutral atoms on the one hand; and CN and OH on the other hand.  \rev{CH is associated with both families, but with the largest CH column densities usually occurring in gas with a high molecular fraction}. Because of the effects of dust extinction, optical/UV observations of absorption by molecules
in the Galactic disk are limited to relatively nearby material, in contrast to the case of the submillimeter observations described in Section 3.3 below.  At centimeter-wavelengths, two hydride molecules -- OH \citep[e.g.][and references therein]{allen:15} and CH \citep[e.g.][]{chastain:10} -- have been widely observed in diffuse interstellar gas clouds, by means of their lambda-doubling transitions at 18 cm and 9 cm respectively.  Such transitions, which are typically observed in emission, have provided a valuable tracer of molecular hydrogen, 
as discussed in Section \ref{sec:diagnostic} below.

\begin{table}
\tabcolsep7.5pt
\caption{Typical hydride abundances in diffuse clouds}
\label{abundances}
\begin{center}
\begin{tabular}{@{}l|c|c|l|l@{}}
\hline
Molecule & Average abundance          & Average abundance    & Method &Reference for hydride\\
        & relative to H \re{or H$_2$} & (fraction of gas     &         & column densities\\ 
       &                              & phase elemental$^a$) &     & \\ 
\hline
CH & \re{$3.5 \times 10^{-8}$} 		& $1.3 \times 10^{-4}$ & V$^b$, UV$^c$, M$^d$ & \cite{sheffer:08}\\
CH$_2$ & \re{$1.6 \times 10^{-8}$} 	& $6   \times 10^{-5}$ & S, CH, M & \cite{polehampton:05}\\
CH$^+$ & $6 \times 10^{-9}$  		& $4   \times 10^{-5}$ & V, Re, M & \cite{gredel:97,crane:95}\\
CH$_3^+$ & $< 2 \times 10^{-9}$  		& $< 1.4   \times 10^{-5}$ &  Re, V & \cite{indriolo:10}\\
\hline
 NH & \re{$8 \times 10^{-9}$}  		& $6 \times 10^{-5}$ & S, CH, 1 & \cite{persson:10}\\
 NH$_2$ & \re{$4 \times 10^{-9}$}  	& $3   \times 10^{-5}$ & S, CH, 1 & \cite{persson:10}\\
 NH$_3$ & \re{$4 \times 10^{-9}$}  	& $3   \times 10^{-5}$ & S, CH, 1 & \cite{persson:10}\\
NH$^+$ & $< 4 \times 10^{-10}$           &   $< 6   \times 10^{-6}$ & S, HI, 1        & \cite{persson:10}\\
\hline
OH & \re{$1 \times 10^{-7}$}		& $1.6   \times 10^{-4}$ & V, S, CH, M & \cite{wiesemeyer:15,lucas:96}\\
H$_2$O & \re{$2.4 \times 10^{-8}$}	& $4    \times 10^{-5}$ & S, CH, M & \cite{flagey:13}\\
OH$^+$ & $1.2 \times 10^{-8}$ 		& $4   \times 10^{-5}$ & S, HI, M$^d$ & \cite{indriolo:15}\\
H$_2$O$^+$  & $2 \times 10^{-9}$	        & $6.5 \times 10^{-6}$ & S, HI, M$^d$ & \cite{indriolo:15}\\
 H$_3$O$^+$  & \re{$2.5\times 10^{-9}$}  & $4   \times 10^{-6}$ & S, CH, 1 & \cite{lis:14}\\
\hline
HF & \re{$1.4 \times 10^{-8}$} 		&  	0.4		& S, CH, M & \cite{sonnentrucker:10}\\
\hline
SH & \re{$1.1 \times 10^{-8}$} 		& $4   \times 10^{-4}$ & S, CH, M & \cite{neufeld:15a}\\
H$_2$S & \re{$5 \times 10^{-9}$} 	& $1.8 \times 10^{-4}$ & S, CH, M & \cite{neufeld:15a}\\
SH$^+$ & $1.1 \times 10^{-8}$		& $8   \times 10^{-4}$ & S, HI, M & \cite{godard:12}\\
\hline
HCl &  \re{$1.5 \times 10^{-9}$}	& 0.004 & S, CH, 1 & \cite{monje:13}\\
HCl$^+$ & 	$8 \times 10^{-9}$	& 0.04 & S, CH, M & \cite{deluca:12}\\
H$_2$Cl$^+$ &$3 \times 10^{-9}$		& 0.02  & S, CH, M & \cite{neufeld:15b}\\
\hline
ArH$^+$ & $3 \times 10^{-10}$ 		& $1 \times 10^{-4}$ & S, HI, M$^e$ & \cite{schilke:14}\\ 
\hline

\end{tabular}
\end{center}
\begin{tabnote}
$^{\rm a}$Assumes the following gas-phase elemental abundances relative to hydrogen in the diffuse ISM: carbon = $1.4 \times 10^{-4}$; oxygen = $3.1 \times 10^{-4}$; nitrogen = $6.8 \times 10^{-5}$; fluorine = $1.8 \times 10^{-8}$; sulphur = $1.4 \times 10^{-5}$; chlorine = $1.8 \times 10^{-7}$; argon = $3.2 \times 10^{-6}$

$^{\rm b}$Method for determination of hydride column density: V = visible or infrared observations of nearby stars;
S = submillimeter, FIR (CH$_2$) or millimeter (H$_2$S) wavelength observations. Abundances determined from visible observations refer to the solar neighborhood while abundances determined from (sub)millimeter and FIR observations also probe the 
Galactic plane  halfway between the Galactic center and the solar circle

$^{\rm c}$Method for determination of H$_2$ or atomic hydrogen column density: 
UV = UV observations of H$_2$ Lyman and Werner bands; 

Re = reddening of background star; CH = submillimeter observations of CH, for an 
assumed $N({\rm CH})/N({\rm H}_2)$ ratio of $3.5 \times 10^{-8}$; 

HI = H 21 cm absorption line observations

$^{\rm d}$Sample size: M = average (median) over multiple sources; 1 = average (mean) over line-of-sight to W31C
 
$^{\rm e}$Median value excludes clouds in the Galactic Center, where the abundances are typically
an order-of-magnitude higher

\end{tabnote}
\label{table:abundances}
\end{table}

\subsection{FIR and submillimeter observations of absorption by interstellar hydrides }

\begin{marginnote}[
]
\entry{{\it Herschel}}{
The {\it Herschel} Space Observatory, a FIR and submillimeter satellite observatory  (55 - 672$\mu$m) with a 3.5\,m primary mirror, operated by  ESA between 2009 and 2013, and with important participation from NASA.}
\end{marginnote}

Over the past five years, {\it Herschel} and SOFIA have provided unique opportunities to observe submillimeter absorption by many interstellar hydrides in diffuse clouds along the sight-lines to bright submillimeter continuum sources.  Heterodyne spectrometers available on both these observatories -- the HIFI instrument on {\it Herschel} and the GREAT instrument on SOFIA -- have provided extremely high (sub-km/s) spectral resolution that is unmatched by other spectrometric techniques operating at submillimeter, infrared or visible wavelengths.   {\it Herschel} observations have led to the discovery of interstellar SH$^+$, HCl$^+$, H$_2$Cl$^+$, H$_2$O$^+$ and ArH$^+$, and the first extensive observations of HF and OH$^+$; SOFIA observations have enabled the first detection of interstellar SH, and the first heterodyne observations of the 2.5 THz ground state rotational line of OH.

\begin{marginnote}[
]
\entry{SOFIA}{
The Stratospheric Observatory For Infrared Astronomy is an airborne 2.7\,m telescope, covering the wavelength range 1 - 655~$\mu$m, operated by NASA and the German Aerospace Center (DLR).}
\end{marginnote}

\begin{figure*}
\includegraphics[width=5 in, angle=0]{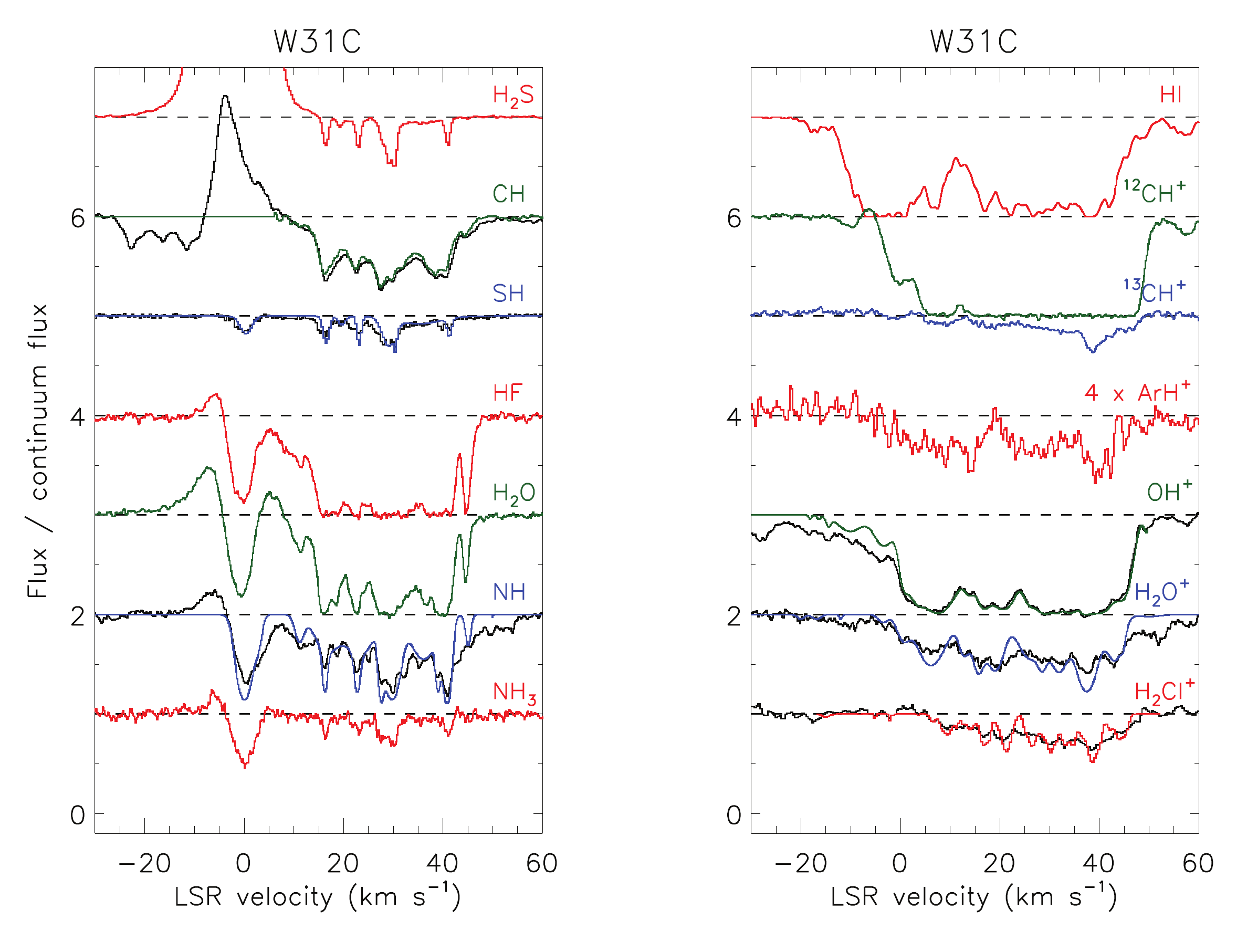}
\caption{Absorption spectra observed toward W31C.  Left panel from top to bottom: 
spectra of H$_2$S $1_{10}-1_{01}$ at 168.8 GHz (based on data published by \cite{neufeld:15b});
CH $\,^2\Pi_{3/2}\,J=3/2-^2\Pi_{1/2}\,J=1/2$ at 532.7/536.8 GHz \citep{gerin:10a}; 
SH $\,^2\Pi_{3/2}J=5/2 - 3/2$ at 1381~GHz \cite{neufeld:15b}; HF $J=1-0$ at 1232~GHz;
H$_2$O $1_{11}-0_{00}$ at 1113~GHz \citep{neufeld:10};  
NH $J_N = 2_1-1_0$ at 974.5~GHz; NH$_3$ $J_K = 2_1-1_1$ at 
1215.2~GHz \citep{persson:10}.  Right panel from top to bottom:
H{\sc{i}} 21 cm (Winkel et al.\, in preparation); $\rm ^{12}CH^+$ $J=1-0$ at 835.1~GHz;
$\rm ^{13}CH^+$ $J=1-0$ at 830.2~GHz \citep{godard:12}; $\rm ArH^+$ $J=1-0$ at 617.5~GHz 
\citep{schilke:14}; OH$^+$ $^2\Pi_{3/2}J=5/2 - 3/2$ at 971.8~GHz; H$_2$O$^+$ 
$1_{11}-0_{00}$ at 1115~GHz \citep{gerin:10}; H$_2$Cl$^+$ $1_{11}-0_{00}$ at 485.4~GHz
\citep{neufeld:12}.  For species with a partially-resolved hyperfine splitting 
(CH, SH, NH, OH$^+$, H$_2$O$^+$), black histograms show the observations and color histograms
show the hyperfine-deconvolved spectra.  For clarity, the spectra are separated by 
vertical offsets.  The ArH$^+$ spectrum is expanded by a factor 4 (in addition to being 
translated) so that the relatively weak absorption is clearer.}
\label{fig:absorption_spectra}
\end{figure*}

Submillimeter absorption-line studies with {\it Herschel} and SOFIA have typically made use of strong continuum sources in the Galactic plane, at distances up to $\sim 11$~kpc from the Sun and with sightlines that can intersect multiple diffuse clouds in foreground spiral arms.   
Thanks to the differential rotation of the Galaxy, different diffuse clouds along a single sightline give rise to separate velocity components in the observed absorption spectra.  {\it Herschel} observations have been performed towards $\sim 20$ such continuum sources, revealing $\sim 100$ distinct components in which molecular column densities can be determined.  

\begin{figure}
\includegraphics[width=5 in, angle=0]{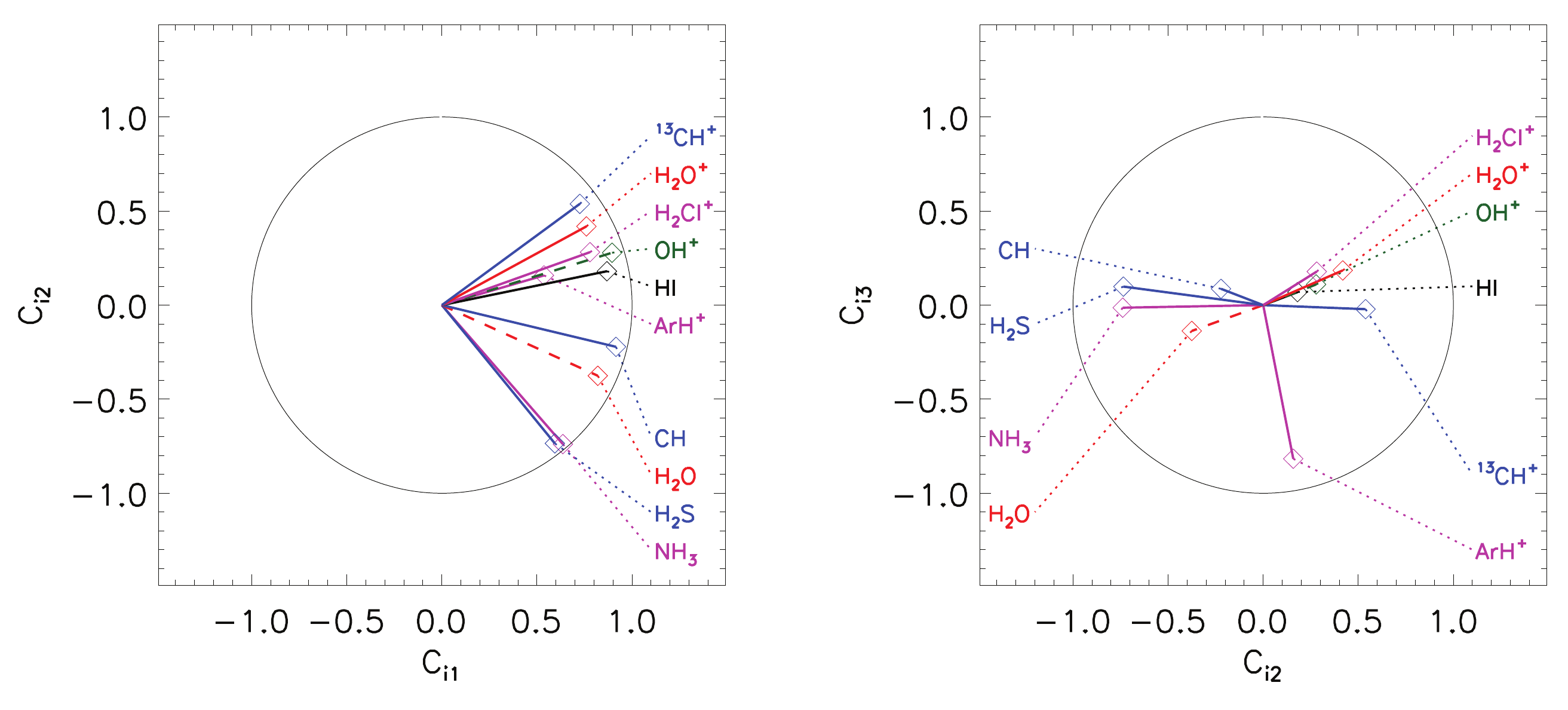}
\caption{Results of PCA analysis for various species observed toward W31C, modified from \citet{neufeld:15a}.}
\label{fig:pca}
\end{figure}

\textbf{Figure \ref{fig:absorption_spectra}} shows the spectra of a dozen hydride molecules observed along the sightline to W31C, a region of high-mass star formation located 4.95 kpc from the Sun, along with the H{\sc{i}} 21 cm absorption line.  All the spectra shown in Figure \ref{fig:absorption_spectra} show clear evidence for absorption by foreground material in diffuse clouds along the sight-line.  For some molecules (e.g. SH and $\rm H_2$S), the observed absorption is concentrated into five or six narrow features with linewidths $\sim 1 - 3\rm \,km\,s^{-1}$; other molecules -- including most of the molecular ions -- show a more extended distribution in velocity space and are less strongly concentrated into narrow absorption components.  \cite{neufeld:15a} have presented a more quantitative examination of the similarities and differences between these absorption spectra with the use of Principal Component Analysis (PCA).  Here, each optical depth spectrum (i.e. plot of optical depth against velocity) is represented as a linear combination of a set of orthogonal (i.e. uncorrelated) eigenfunctions, $S_j$, the latter chosen such that the first eigenfunction contains as much of the variation as possible, the second eigenfunction containing as much of the remaining variation as possible, and so forth.  Typically, the spectrum of any species, $i$, may be represented by a linear combination of the first two or three eigenfunctions, $C_{i1}S_1 + C_{i2}S_2\,(\, + C_{i3}S_3\,)$,  which
provides an excellent fit to the observed optical depth spectra, and the {\it coefficients} within that linear combination -- $C_{i1}$, $C_{i2}$, and $C_{i3}$ -- can be used to describe the distribution of each absorbing species.  \textbf{Figure \ref{fig:pca}} shows the first three coefficients obtained from a joint analysis of ten species along the W31C sightlines.  Here, diamonds indicate the coefficients obtained for each species in the $C_{i1}$ - $C_{i2}$ (left panel) and $C_{i2}$ - $C_{i3}$ (right) planes, and solid lines connect the origin to each diamond.  In the limit where the {\it first two} principal components account for most of the variation in the spectra, the diamonds in the left panel lie on the unit circle, as explained by \cite{neufeld:15a}; moreover, the angle between the lines corresponding to any pair of species in that limit is equal to the arccosine of their correlation coefficient.  \rev{The PCA results confirm that most of the molecular ions show a common
distribution, which is similar to that of atomic hydrogen, while most
neutral species - particularly NH$_3$ and H$_2$S - tend to follow a distribution
that is uncorrelated with H{\sc{i}}.}
The ArH$^+$ molecular ion appears largely uncorrelated both with the other molecular ions and with the neutral molecules.

In Table \ref{table:abundances}, we summarize the typical hydride abundances derived for the diffuse ISM.  In the case of molecular ions (with the exception of H$_3$O$^+$), these are shown relative to atomic hydrogen, whereas for neutral molecules, they are shown relative to H$_2$.  In both cases, the corresponding abundances are also given relative to the gas-phase abundance of the relevant heavy element.  The values shown in Table~\ref{table:abundances} indicate that most hydride molecules in diffuse gas are only minor reservoirs of the heavy element they contain, with most heavy elements remaining overwhelmingly in atomic or singly-ionized form.  The notable exception is HF, which, in diffuse regions with a large molecular hydrogen fraction, can account for most of the gas-phase fluorine.  Chlorine is another element with a fairly strong tendency toward hydride formation; in clouds with a modest H$_2$ fraction, H$_2$Cl$^+$ can account for several percent of the gas-phase chlorine abundance.  These behaviors are \rev{thought}  to reflect the thermochemical considerations summarized in Figure \ref{fig:thermochem}; fluorine and chlorine are the only elements that can react exothermically with H$_2$ when in their dominant stage of ionization (i.e.\ F and Cl$^+$) in diffuse clouds.

\section{HYDRIDES IN DENSE, STRONGLY UV-IRRADIATED GAS} \label{sec:pdr}

While many aspects of diffuse molecular clouds  can be understood in the framework of low $A_V$,
low FUV field PDR models \citep[for a review see][]{hollenbach:99}, in this section we focus
on the presence of hydrides in more strongly UV-irradiated ($\chi$$>$10$^3$ times the average
interstellar radiation field), and denser gas ($n_{\rm H}$$>$10$^{4}$\,cm$^{-3}$). These are
conditions typical of the \textit{surface} of dense molecular clouds illuminated by nearby
massive stars, and also of the outer layers of irradiated protoplanetary disks (e.g.
proplyds). Such ``dense PDRs'' are regulated by the presence of stellar FUV photons up to
several magnitudes of extinction (high $A_V$). UV photons with energies higher than 13.6\,eV
 ionize  hydrogen atoms but they are absorbed in the adjacent  H\,{\sc ii} regions. FUV
photons, however, do dissociate molecules and  ionize molecules and atoms with ionization
potentials below 13.6\,eV (C, S, Si, Na, Fe, etc.). Therefore, like  diffuse clouds, dense
PDRs are predominantly neutral, with maximum ionization fractions of $n_{\rm e}$/$n_{\rm
H}$$\simeq$10$^{-4}$, roughly the gas-phase carbon abundance.
The chemistry of dense PDRs is driven by photo-reactions \citep[][]{Sternberg:95} and is
also characterized by the warm temperatures attained by the gas ($T$$\simeq$100-1000\,K) and
by the presence of large fractions of (FUV-pumped) vibrationally excited molecular hydrogen,
H$_{2}^{*}$ \citep[e.g.][]{Black:87}.

Despite the large FUV radiation fields, H$_2$ molecules form on the surfaces of relatively
warm dust grains \citep[$T_{\rm dust}$$\gtrsim$50\,K for $\chi$$\gtrsim$10$^3$, e.g.][and
references therein]{cazaux:04,bron:14}. As the H$_2$ abundance increases from the PDR edge to the
cloud interior, reactions of H$_2$ with neutral and singly ionized atoms initiate the
chemistry and  lead to the formation of hydrides. Owing to the higher gas and electron
densities  compared to diffuse clouds, the excitation temperatures rise above the background
temperature and rotational line \textit{emission} is observed. Recent detections in dense
PDRs include NH$_3$, CH$^+$, SH$^+$, OH$^+$, H$_2$O, OH, HF, CH and H$_2$Cl$^+$
\citep[e.g.][]{Batrla:03,Habart:10,Goicoechea:11,nagy:13,van_der_Tak:12,van_der_Tak:13,
neufeld:12}.

\subsection{The role of vibrationally excited H$_2$ in the chemistry} \label{sub-sec:vibH2}

The absorption of FUV photons in the Lyman \mbox{($B\,^1\Sigma_u-X\,^1\Sigma_{g}^{+}$)} and
Werner \mbox{($C\,^1\Pi_u-X\,^1\Sigma_{g}^{+}$)} bands produces molecular hydrogen in
excited electronic states. 10\%~of absorptions are followed by emission into the unbound
continuum, leading to dissociation of the molecule.  90\%~of the absorptions, however, are
followed by rapid  radiative decay to the ground
electronic state~$X$. See an early historical review by \citet{field:66} and later
quantitative developments by e.g.~\citet[][and references therein]{black:76}. PDRs thus
naturally produce large fractional abundances,
\mbox{$f^*$=$n$(H$_{2}$\,$v$$\geq$1)/$n$(H$_{2}$\,$v$$=$0)}, of vibrationally excited molecular hydrogen, H$_{2}^{*}$.
In low density PDRs ($\lesssim$10$^4$\,cm$^{-3}$) H$_2$ is excited by FUV absorption
followed by radiative fluorescent emission. The main gas heating mechanism is the
photoelectric effect on PAHs and grain surfaces. 
For higher gas
densities, collisional de-excitation through inelastic H-H$_2$ and H$_2$-H$_2$ collisions
start to be important, and collisional de-excitation of FUV-pumped H$_2$ can dominate the
gas heating. In addition, pure rotational lines of HD \citep{Wright:99} and H$_2$
\citep[][]{Allers:05} can been detected. Their  low critical densities make them good
diagnostics of the warm gas temperature.

The H$_2$ $v$$=$1 levels  have energies $>$0.5\,eV ($\sim$5800\,K) above the   ground state
$v$$=$0 and  this energy can be used to overcome the endothermicity (or energy barrier) of
chemical reactions involved in the formation of  some hydrides, thus enhancing their
production rate in PDRs. A very favorable case is CH$^+$. As previously noted, the
\mbox{C$^+$\,+\,H$_2$\,$\rightarrow$\,CH$^+$\,+\,H} reaction is endothermic by $\sim$4300\,K
(see Figure~\ref{fig:thermochem}), 
temperatures ($T$$>$1,000 K) are required to produce significant amounts of CH$^+$ when
H$_{2}$ is in the $v$$=$0 state. \mbox{Laboratory} experiments \citep{Hierl:97} and  quantum
calculations \citep[e.g.][]{zanchet:13}, however, show that the reaction becomes
exothermic when  H$_{2}^{*}$, and not H$_{2}$($v$$=$0),  is the reactant. The reaction rate
$k^*$ then reaches the Langevin limiting case, and becomes almost independent of the
temperature. In practice, enhanced XH$^{+}$ (or XH) hydride abundances will be produced
through reactions \mbox{X$^{+}$\,+\,H$_{2}^{*}$\,$\rightarrow$\,XH$^{+}$\,+\,H} (or
\mbox{X\,+\,H$_{2}^{*}$\,$\rightarrow$\,XH\,+\,H}) if the product \mbox{$f^* \Delta k\,(T)$}
is larger than~1. Here, \mbox{$\Delta k\,(T)=k^*\,(T)/k^0\,(T)$} is the reaction rate ratio
when  H$_{2}^{*}$ and H$_{2}$($v$=0) are the reactants  respectively
\citep[see][]{Agundez:10}. In addition, $k^*\,(T)$, the specific reaction rate with
H$_{2}^{*}$,  has to be large enough compared to other reactions removing the atom X$^{+}$
(or X), for example, X$^{+}$ neutralization reactions with PAHs and small grains.

\begin{figure}[t] \includegraphics[scale=0.7,angle=0]{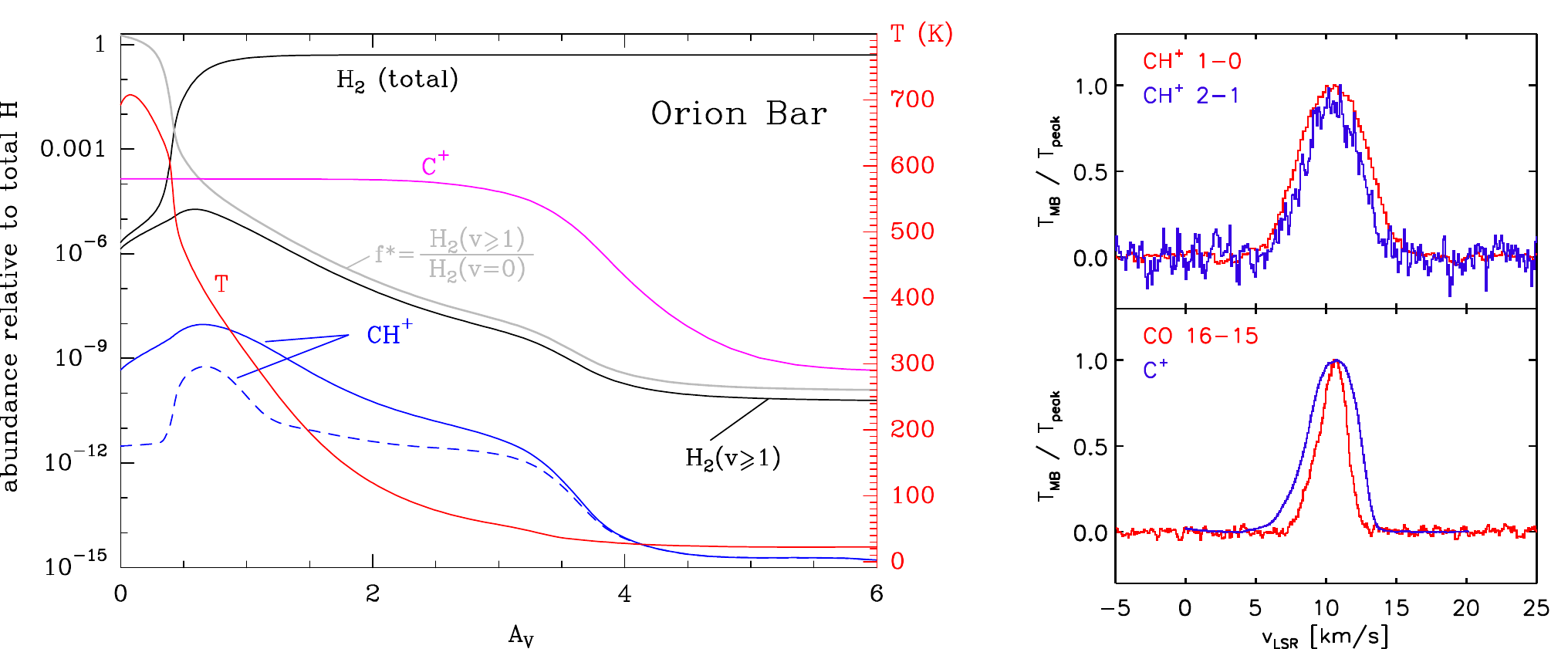}
\caption{(Left): PDR model adapted to the physical conditions in the Orion Bar: $n_{\rm
H}$$=$10$^5$\,cm$^{-3}$ and $\chi$$=$3$\times$10$^4$. Abundances relative to $n_{\rm H}$ are
shown  for H$_{2}$, H$_{2}$(v$\geq$1)$=$H$_{2}^{*}$, C$^{+}$, and CH$^+$ as a function of
depth $A_V$ into the cloud.  Also shown is
\mbox{$f^*$=$n$(H$_{2}$\,$v$$\geq$1)/$n$(H$_{2}$\,$v$$=$0)}, the fraction of vibrationally
excited H$_2$ with respect  to the ground vibrational state, 
 and the gas temperature $T$. For the CH$^+$ abundance profile, the dashed
line is for a specific model that only considers reactions of C$^+$  with H$_2$ in the
ground  vibrationally state (v$=$0). The continuous curve shows the CH$^+$ abundance
including also state-to-state reactions of C$^+$  with H$_{2}^{*}$
\citep[adapted from][]{Agundez:10}. (Right): Velocity-resolved observations of the Orion Bar with
\textit{Herschel}/HIFI. CH$^+$ $J$=1-0 and 2-1 lines are broader than those of [C{\sc ii}]
and CO $J$=16-15. This may be the signature of  formation pumping as explained at the end of Sect. \ref{sub_sec:formpump} \citep[from][]{nagy:13}.}
\label{fig_pdr_chp} \end{figure}

Rotationally excited CH$^+$ emission lines were first detected toward planetary nebula
NGC\,7027 \citep{Cernicharo:97}. These lines arise from the warm and dense PDR surrounding
the ionized nebula that is  irradiated by the strong UV field from the hot central star
($T_{\rm eff}$$\approx$200,000~K). The same pure rotation emission lines have been  detected
in the Orion Bar \citep[][see \textbf{Figure~\ref{fig_pdr_chp} right}]{nagy:13} and in the
irradiated surface of the protoplanetary disk around Herbig~Be star HD\,100546
\citep{thi:11}. As there is no reaction pathway that can efficiently produce CH$^+$ in
FUV-shielded cold gas, CH$^+$ is expected to peak near the PDR edge where  C$^+$ and
H$_{2}^{*}$ are abundant (see \textbf{Figure~\ref{fig_pdr_chp} left}). A similar argument
applies to SH$^+$, with the  difference that the 
\mbox{S$^+$\,+\,H$_2$\,$\rightarrow$\,SH$^+$\,+\,H} reaction is even more endothermic
($\sim$10,000\,K) and only becomes exothermic when H$_{2}^{*}$ is in the $v$=2 or higher
vibrational states \citep{Zanchet:13b}. Line emission from both ions has been observed in
the Orion Bar   with \textit{Herschel} \citep[][]{nagy:13}, but $^{13}$CH$^+$ and SH$^+$ can
be detected from ground-based telescopes such as ALMA. Note that rotationally excited CH$^+$
has also been recently observed in visible absorption \citep[][]{oka:13}  toward lower
density PDR environments close to hot massive stars, where vibrationally excited H$_2$ is
also observed in FUV absorption \citep{rachford:14}.

\begin{marginnote}[
]
\entry{ALMA}{ The Atacama Large Millimeter/submillimeter Array is an ensemble of 66
(sub)millimeter telescopes  operating between 80 and 900~GHz. 
}
\end{marginnote}

The role of FUV-pumped H$_{2}^{*}$ in the chemistry may also be important at the scales of
entire molecular cloud complexes. In template star-forming regions like Orion, $>$98$\%$ of
the large-scale near-IR  H$_2$ emission is thought to arise from FUV-excited fluorescent
emission in  PDR gas \citep{Luhman:94}. This widespread H$_{2}^{*}$ emission traces the
extended molecular gas that is illuminated by the  strong FUV radiation field from massive
stars in the Trapezium cluster. Its presence likely explains the large-scale CH$^+$ $J$=1-0
emission detected in Orion molecular cloud  (J.R. Goicoechea priv.comm.). The high
abundances of  H$_{2}^{*}$ in local massive star-forming regions may have consequences for 
the interpretation of hydride spectra in the extragalactic context, as many actively star-forming
galaxies are dominated by  PDR emission. In  diffuse molecular gas (much lower density and
FUV fluxes),  the H$_{2}^{*}$ abundance and $f^*$ fractions are significantly
lower. Indeed, the large CH$^+$ and SH$^+$  column densities inferred from line absorption
observations of  diffuse clouds cannot be explained by standard PDR models adapted to
the $n_{\rm H}$$\simeq$100\,cm$^{-3}$  and $\chi$$\simeq$1 conditions prevailing in these
clouds \rev{\citep[e.g.][and references to older models therein]{godard:12}}.

\subsection{Oxygen hydrides}

\begin{figure}[t] \includegraphics[scale=0.75,angle=0]{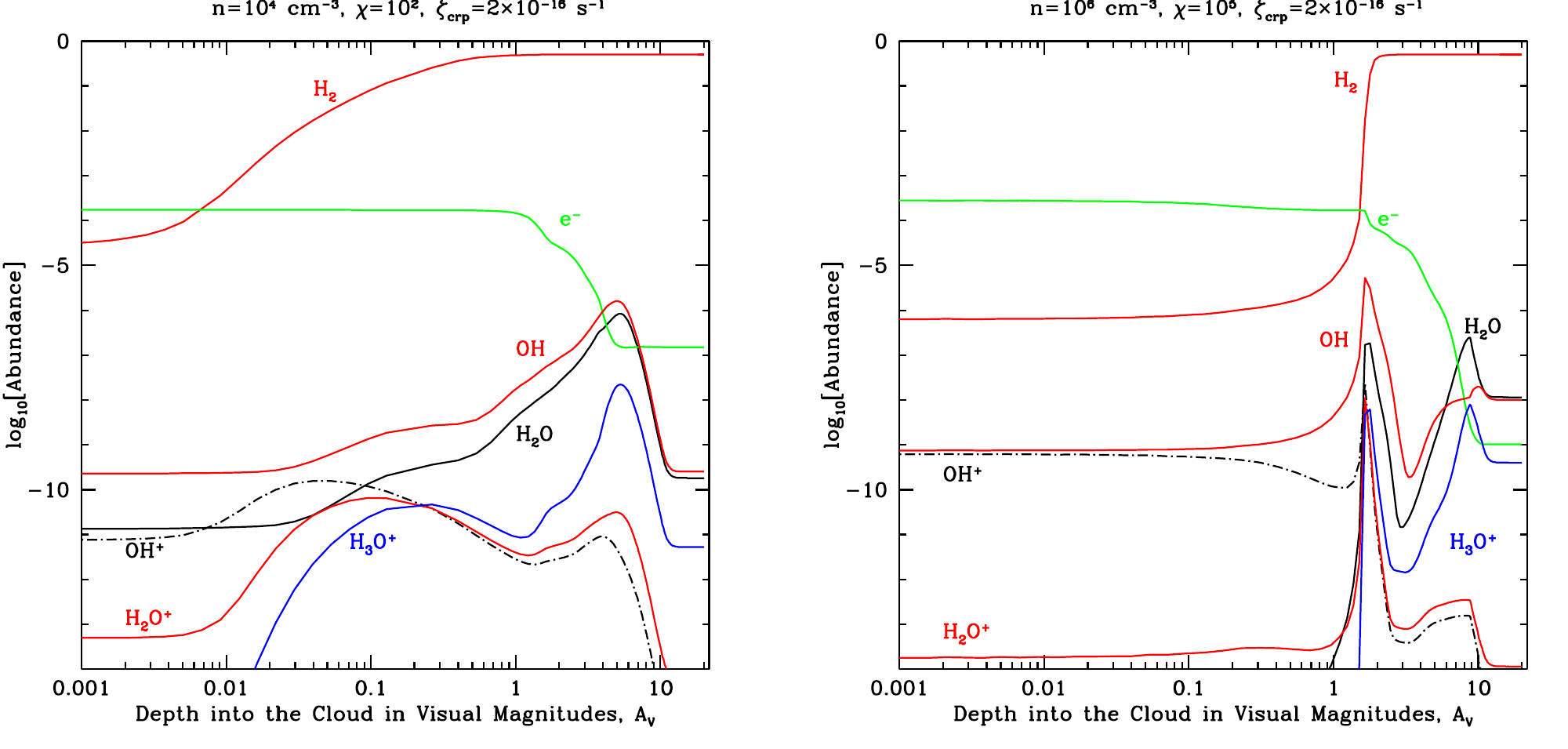}
\vspace{0.1cm} \caption{Predicted gas-phase abundances of several hydrides with respect to H
nuclei as a function of depth $A_V$ into the cloud for PDRs with $n_{\rm
H}$=10$^4$\,cm$^{-3}$, $\chi$=100, and $\zeta_{\rm CR}$=2$\times$10$^{-16}$\,s$^{-1}$
(left), and with $n_{\rm H}$=10$^6$\,cm$^{-3}$ and $\chi$=10$^5$ (right). From
\citet[][]{hollenbach:12}.} \label{fig_pdr_mod_hol} \end{figure}

Relatively low water vapor abundances are expected in dense PDRs. H$_2$O has a high
adsorption energy and thus  it freezes at $T_{\rm dust}$$<$100\,K in dense cloud conditions
(as compared to $\sim$20\,K for CO). Photodesorption of water ice mantles takes place in
PDRs but H$_2$O molecules are readily photodissociated by the intense FUV radiation field
\citep{hollenbach:09}. This process results predominantly in OH\,+\,H 
\citep[e.g.][hereafter vD13]{vandishoeck:13}. Low excitation H$_2$O rotational lines have
been observed toward high FUV flux PDRs such as the Orion Bar \citep{Habart:10,Choi:14} and
Mon~R2 \citep{Pilleri:12}. The inferred gas-phase H$_2$O abundances are below a few
10$^{-7}$. Rotationally excited OH emission lines have been also detected in the Orion Bar.
They  are consistent with a \mbox{OH/H$_2$O$>$1} column density ratio \citep{Goicoechea:11},
similar to the diffuse cloud values, whereas \mbox{OH/H$_2$O$<$1} ratios are typically
observed toward warm shocked gas (see next Section). Regarding OH formation, the dominant
route near the edge of a strongly UV-irradiated PDR is the  endothermic reaction
\mbox{O\,+\,H$_2$\,$\rightarrow$\,OH\,+\,H} (see the first OH abundance peak in
\textbf{Figure~\ref{fig_pdr_mod_hol} right}). \rev{However, the reaction remains slow when the H$_2$ is vibrationally 
excited \citep{Sultanov:05}}. Consequently, the OH formation
rate is likely more sensitive to the gas temperature  than to the  H$_{2}^{*}$ abundance.
For lower FUV fields and temperatures, OH forms from H$_3$O$^+$ electron recombination and
from grain surface reactions \citep[e.g.][]{hollenbach:09,hollenbach:12}.

In the Milky Way, OH$^+$ rotational line \textit{emission} has been detected  toward high
electron density environments: the Orion Bar \citep{van_der_Tak:13}, planetary nebulae
hosting hot central stars with $T_{\rm eff}$$>$100,000~K \citep{Aleman:14}, and toward the
Supernova remnant in the Crab Nebula \citep{barlow:13}. The non-detection of  H$_2$O$^+$ and
H$_3$O$^+$ emission in these environments indicates that OH$^+$ lines arise in  gas layers
where most hydrogen is in atomic form, and OH$^+$ is predominantly formed by the
\mbox{O$^+$\,+\,H$_2$\,$\rightarrow$\,OH$^+$\,+\,H} exothermic reaction. 
 Figure~\ref{fig_pdr_mod_hol} left shows predictions of a steady-state
PDR model for a $n_{\rm H}$=10$^4$\,cm$^{-3}$ cloud illuminated by a FUV field that is 100 times
the mean ISRF. These conditions are representative of a molecular cloud  close to massive O
and B stars. The model includes reactions with PAHs and freeze-out of gas-phase species
(important at $A_V$$>$4). It shows that OH$^+$ peaks closer to the PDR edge than  other
oxygen hydrides. Figure~\ref{fig_pdr_mod_hol} right shows the results for a higher density,
$n_{\rm H}$=10$^6$\,cm$^{-3}$, and higher FUV irradiation model, $\chi$$=$10$^5$. These conditions
are more typical of the compressed PDR layers at the boundary of a H\,{\sc ii} region
surrounding a massive star  or the surface layers of protoplanetary disks. In this model, H$_2$ collisional de-excitation
dominates the gas heating and the gas attains higher temperatures ($T$$\simeq$1000\,K at
$A_V$$\simeq$1). This triggers a fast warm gas-phase  in which many
endothermic reactions can proceed rapidly. \revv{The dense PDRs located  at the walls of outflow cavities  irradiated  by a high mass young stellar object have been modeled by \citet{bruderer:10}, who predicted high abundances of CH$^+$, OH$^+$ and NH$^+$ within such regions. }

\subsection{Formation pumping and the excitation of reactive hydride ions} \label{sub_sec:formpump}

Reactive collisions (collisions that lead to a chemical reaction) influence the excitation
of molecules when the time-scale of  the chemical reaction becomes comparable, or shorter,
than that of nonreactive collisions. This makes the lifetime of the molecule so short that
it does not get ``thermalized" by collisions with non reactive species, or by radiative
absorptions of  the background radiation field. This is referred to as ``formation
pumping'', the molecule retains an excited state due to its formation, and a proper
treatment of the molecule excitation requires including chemical formation and destruction
rates in the statistical equilibrium calculation determining the level populations
\citep[e.g.][]{Black:98}. Reactive ions such as CH$^+$ or OH$^+$ are a special kind because
they can be destroyed on  nearly every collision with the most abundant species: H$_2$, H
and $e^-$ (OH$^+$ in particular does not react with H). Detailed  CH$^+$ excitation models show
that  formation pumping dominates the population of CH$^+$ rotationally excited  levels in
PDRs such as the Orion Bar \citep{godard:13}. In addition, the effects of molecule
destruction and molecule formation in excited levels needs to be taken into account in
chemical networks. 
 It is becoming  common to refer to ``state-to-state'' chemistry when the specific
state-dependent reaction rates are included in models.

Reactions of C$^+$ with H$_{2}$($v$=0, $J$$>$7) or with H$_{2}$($v$$\geq$1) are exothermic
and drive the CH$^+$ production in PDRs such as the Orion Bar (see
Section~\ref{sub-sec:vibH2}).  Therefore, CH$^+$ formation results in an excess of energy
equivalent to several thousand K. This energy may be redistributed   into translational
motion of the newly formed CH$^+$ molecules \citep{Black:98}.  CH$^+$ reacts so quickly with
H, H$_2$ and electrons that its translational motions may never thermalize to the gas
kinetic temperature.
The broader CH$^+$ $J$=1-0  emission line widths observed toward the Orion Bar
($\sim$5.5\,km\,s$^{-1}$, corresponding to a temperature of several thousand K) compared to
those of other molecular lines (usually \mbox{$\sim$2-3\,km\,s$^{-1}$}, see
Figure~\ref{fig_pdr_chp} right) has been interpreted as the signature of CH$^+$  formation 
pumping and of non-thermal velocity distributions \citep{nagy:13}. Formation pumping also
influences  OH$^+$ excitation \citep{Gomez:14} and may also explain the  broad  OH$^+$
emission lines    \citep{van_der_Tak:13}.

\section{HYDRIDES IN SHOCKS AND TURBULENT DISSIPATION REGIONS}
\label{sec:shocks}

\subsection{Shocks}
Several astrophysical phenomena -- including protostellar outflows and supernovae -- give
rise to supersonic motions in the interstellar medium, and these motions can drive shock 
waves that heat and compress the gas.  Such shock waves modify both the chemistry of -- 
and the emission from -- interstellar hydrides, greatly enhancing the abundances of 
certain species, populating high-lying rotational states, 
and triggering strong maser amplification in specific transitions of OH and $\rm H_2O$.

The chemical effects of shock waves include the sputtering of dust grains, which releases grain materials into the gas-phase, and the enhancement of gas-phase endothermic reactions that are negligibly slow at the typical temperatures ($T \le 80\,\rm K$) of cold interstellar gas clouds but can become rapid behind shock waves. In dense molecular clouds, where dust grains are coated with icy grain mantles, the sputtering of grain mantles has a profound effect upon the water vapor abundances behind shocks.  The sputtering process has been the subject of several theoretical studies 
\citep{draine:83,jimenez:08,gusdorf:08}, 
which suggest that grain mantles are completely removed in shocks propagating at velocities, $v_s$,  larger than $\sim 20 - 25 \, \rm km\, s^{-1}$,  resulting in the release of water and other volatile grain mantle materials into the gas-phase.  Observationally, the interstellar water vapor abundance measured by satellite observatories such as the {\it Submillimeter Wave Astronomy Satellite} (SWAS), Odin and {\it Herschel} is known to vary by more than three orders of magnitude (see vD13 and references therein), with the largest abundances observed in warm shocked regions associated with protostellar outflows and supernova remnants.  Here, the water vapor abundance can reach several $\times 10^{-4}$ relative to hydrogen nuclei, an abundance comparable to that of carbon monoxide.  However, many determinations of the water vapor abundances in shocked regions (vD13, their Table 4) are considerably smaller and lie below the predictions for shocks capable of sputtering icy grain mantles.  This discrepancy may reflect the fact that many abundance determinations are averages over a mixture of shocks of varying velocity, some of which are too slow to sputter mantles\rev{, or due to UV irradiation of the shocked gas lowering the H$_2$O abundance and enhancing OH \citep{karska:14,goicoechea:15}}.  In a recent study \citep{neufeld:14} of the far-IR line emission detected by {\it Herschel} and {\it Spitzer} from shocked gas in NGC 2071, W28, and 3C391, the relative line intensities for multiple transitions of H$_2$O, CO and H$_2$ were found to be in acceptable agreement with standard theoretical models for nondissociative shocks that predict the complete vaporization of grain mantles in shocks of velocity $\sim 25 \, \rm km\, s^{-1}$, behind which the characteristic gas temperature is $\sim 1300$~K and the H$_2$O/CO ratio is 1.2.

The second chemical effect of shock waves is to selectively increase the abundances of molecules that can be produced rapidly by endothermic reactions that are slow at the temperatures of cold quiescent gas clouds.  Behind shock waves, elevated gas temperatures $\rm \sim few \,\times 10^2$ to $\rm \sim few \,\times 10^3\,K$ can be sufficient to overcome the barriers for several of the endothermic H-atom abstraction reactions listed in Fig. \ref{fig:thermochem}, greatly increasing the predicted abundances of CH$^+$, SH$^+$, CH, SH, OH and H$_2$O. In addition to the effects of the elevated temperatures behind interstellar shock waves, velocity drifts between the ions (which are coupled to the magnetic field) and the neutral species (which are not) can lead to an additional enhancement in the rates of ion-neutral reactions.
Magnetohydrodynamic shock waves\rev{ propagating at $\sim 10$~km\,s$^{-1}$}, in which such velocity drifts are present, have been invoked \citep[e.g.][]{draine:86, desforets:86a, desforets:86b} to explain large discrepancies between the observed abundances of CH$^+$, SH$^+$, and SH in diffuse molecular clouds and the predictions of standard photochemical models for cold ($T \le 80\, \rm K$) diffuse gas. 

\begin{figure}
\includegraphics[width=12cm]{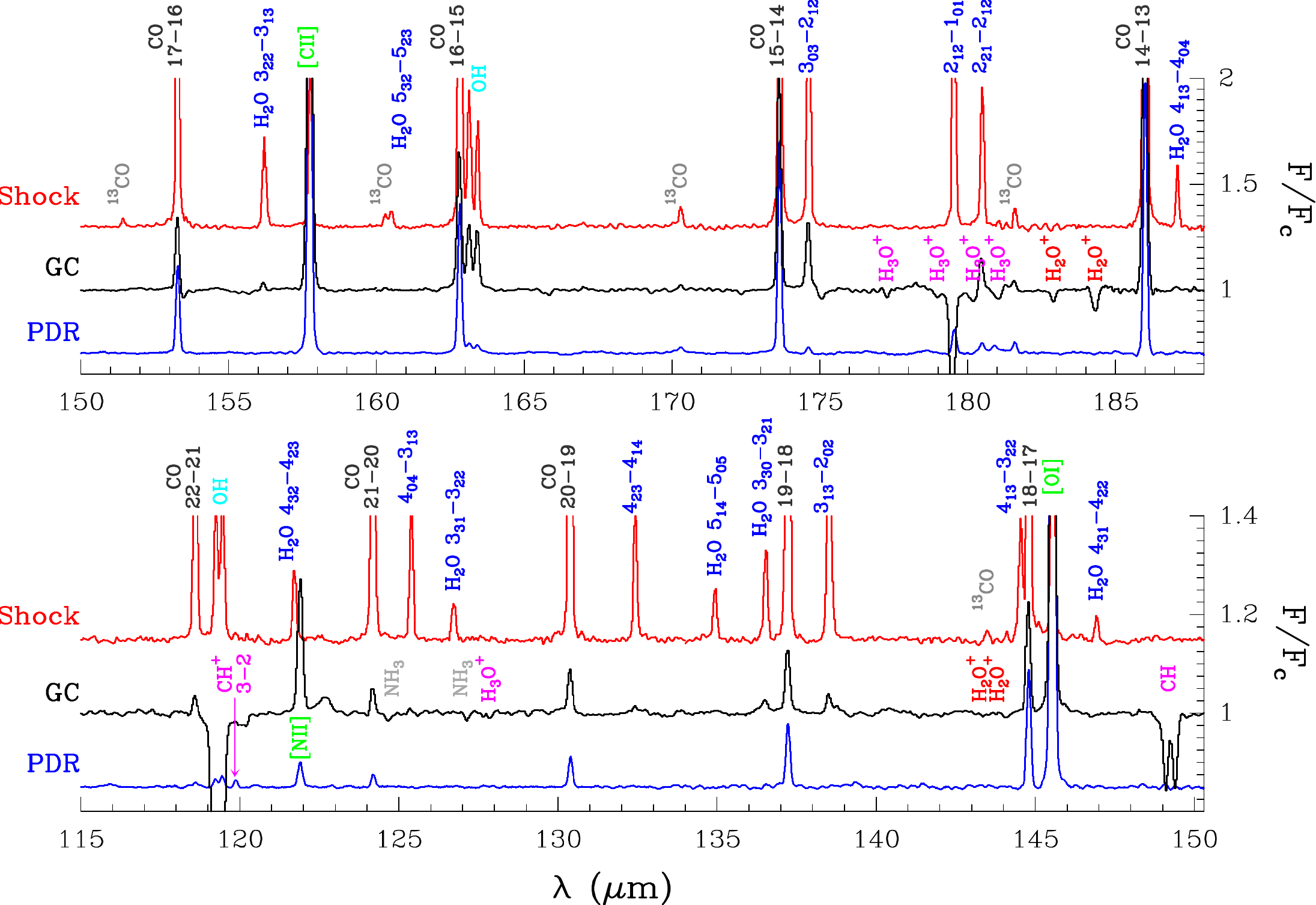}
\caption{Comparative FIR spectra of three template environments in the Milky Way taken with
\textit{Herschel}/PACS  between 115 and 188\,$\mu$m: a molecular outflow and associated shocked gas,  the Galactic Center,
and a highly UV-irradiated PDR. They correspond, from top to bottom, to \mbox{Orion
BN/KL outflows} \citep[red,][]{goicoechea:15}, the circumnuclear disk around
Sgr\,A$^*$ \citep[black,][]{goicoechea:13}, and the Orion Bar PDR (blue, Joblin et
al. in prep.). The ordinate scale refers to the line flux to continuum flux ratio, and
the abscissa to the wavelength in microns. 
The "shock" and "PDR" spectra have been shifted from the F/F$_c$=1 level for
clarity. 
}
\label{fig:pacs}
\end{figure}

In addition to their chemical effects, shock waves propagating in dense molecular gas also populate high-lying rotational states of hydrides that lead to a distinctive emission spectrum dominated by high-lying, collisional-pumped rotational transitions of H$_2$, CO, $\rm H_2O$, and OH \citep{chernoff:82,draine:93}.   {\textbf {Figure \ref{fig:pacs}}} shows a comparison of the
FIR spectra of three environments : a molecular outflow and associated shocked gas,  the Galactic Center combining dense gas and foreground absorption by  diffuse gas, and a highly UV-irradiated dense PDR.  While the shocked region shows intense OH and H$_2$O emission lines, the dense and warm environment of the Galactic Center region has a more diverse spectrum with other hydrides like 
NH$_3$, H$_3$O$^+$ and H$_2$O$^+$ showing up, \rev{indicating the presence of UV and CR irradiated molecular gas. The PDR spectrum is dominated by strong [C{\sc ii}] and [O{\sc i}] emission, with faint lines from  H$_2$O and  rotationally excited CH$^+$.}

\begin{figure}
\includegraphics[width=4.8in]{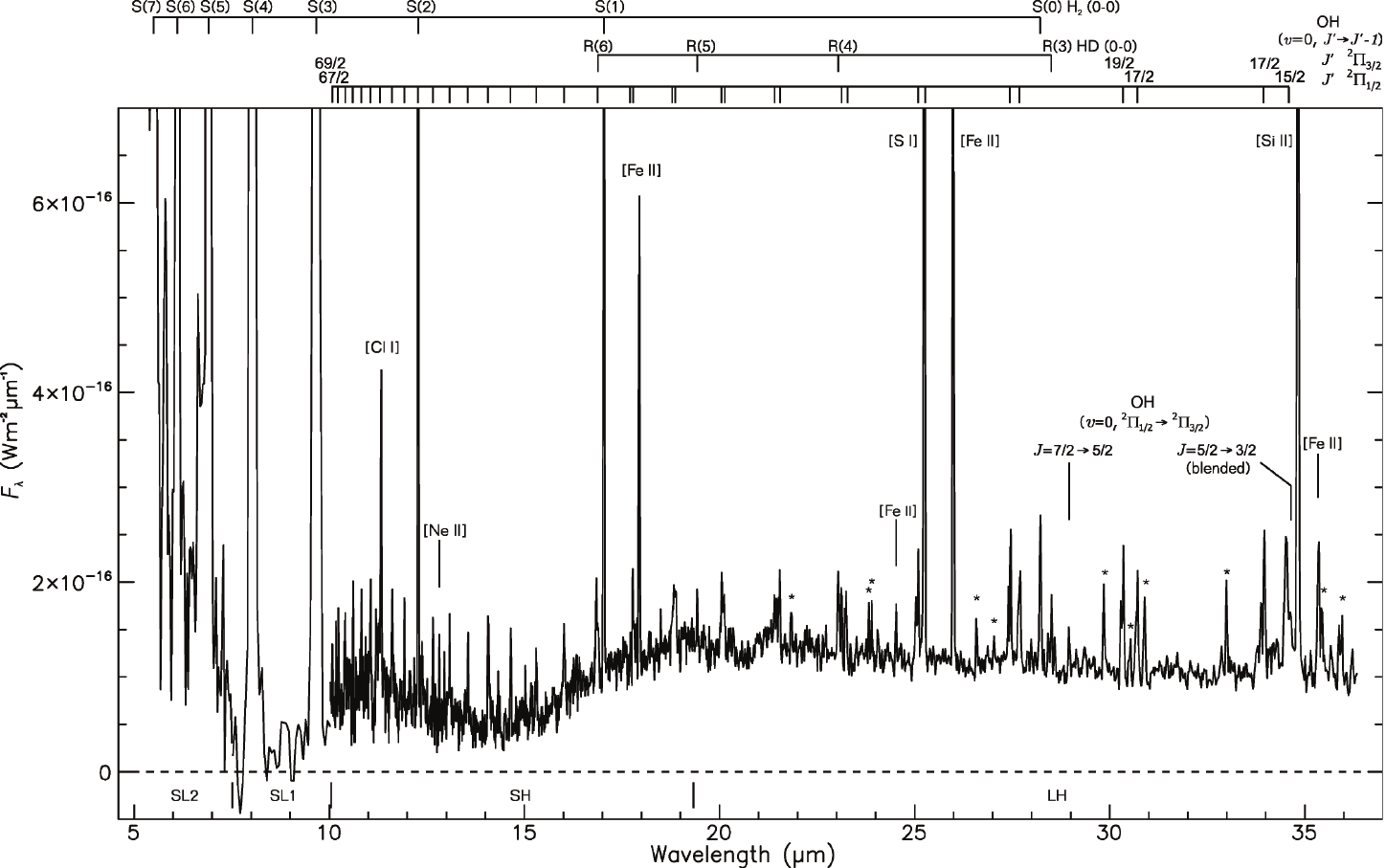}
\caption{{\it Spitzer} spectrum obtained toward HH 211 \citep{tappe:08}. Pure rotational lines of OH, with upper states with $J$ up to 69/2, are clearly detected.  Features marked with an asterisk are pure rotational transitions of H$_2$O.}
\label{fig:T08f2}
\end{figure}

In the case of OH, mid-IR transitions with upper state energies, $E_{\rm u}/k_B$ of several 10$^4$~K can also be excited following the photodissociation of H$_2$O by shock-generated UV radiation; this phenomenon, observed by {\it Spitzer} toward HH 211 \rev{and illustrated in} \textbf{Figure \ref{fig:T08f2} }\citep{tappe:08}, occurs because the photodissociation of H$_2$O following absorption of Lyman alpha radiation leaves OH in states of rotational quantum number 30 - 50.

Along with the normal (non-inverted) OH and H$_2$O line radiation emitted by warm shocked regions, strong maser action is observed in several transitions : these include the 22 GHz $6_{16}-5_{23}$ transition of H$_2$O, frequently observed from outflow-driven shock waves
along with several submillimeter water maser transitions \citep[e.g.][and references therein]{neufeld:13}, and the 1720 MHz transition of OH, 
a signpost of supernova-driven shock waves \citep[e.g][]{frail:98}.
Maser lines are typically characterized by small spots of emission, often unresolved even in long-baseline interferometric observations, narrow line width, and variability on timescales of months or less.  Emission in the 22 GHz water maser transition, in particular, can achieve extraordinarily high brightness temperatures that may exceed 10$^{14}$~K.
Theoretical models for water maser emissions invoke either fast, dissociative jump (``J"-type)
shocks \citep{elitzur:89,hollenbach:13}, or slower non-dissociative ``C-type" shocks in which the fluid velocities vary continuously \citep{kaufman:96}.  Where multiple water maser transitions are observed, maser line ratios can provide constraints on the gas temperature in the masing region and the nature of the shocks that are present.  Paradoxically, non-dissociative shocks can produce hotter {\it molecular} material than faster dissociative shocks, where molecules are quickly destroyed behind the shock front and are reformed only after the gas has cooled to $\sim 500$~K.  In some maser sources, gas temperatures of at least $\sim 900$~K are implied by the observed water maser line ratios, indicating the presence of non-dissociative shocks \citep{melnick:93}.  
Where supernova-driven shock waves encounter molecular clouds, 
widespread 1720 MHz OH maser emission is observed; here, the enhanced column densities of OH needed to explain the observed maser emission are believed to result from the effects of X-rays and/or cosmic rays in dissociating H$_2$O to form OH \citep{wardle:02}.

\subsection{Turbulent dissipation regions}

\begin{figure}
\resizebox{5cm}{!}{%
\includegraphics{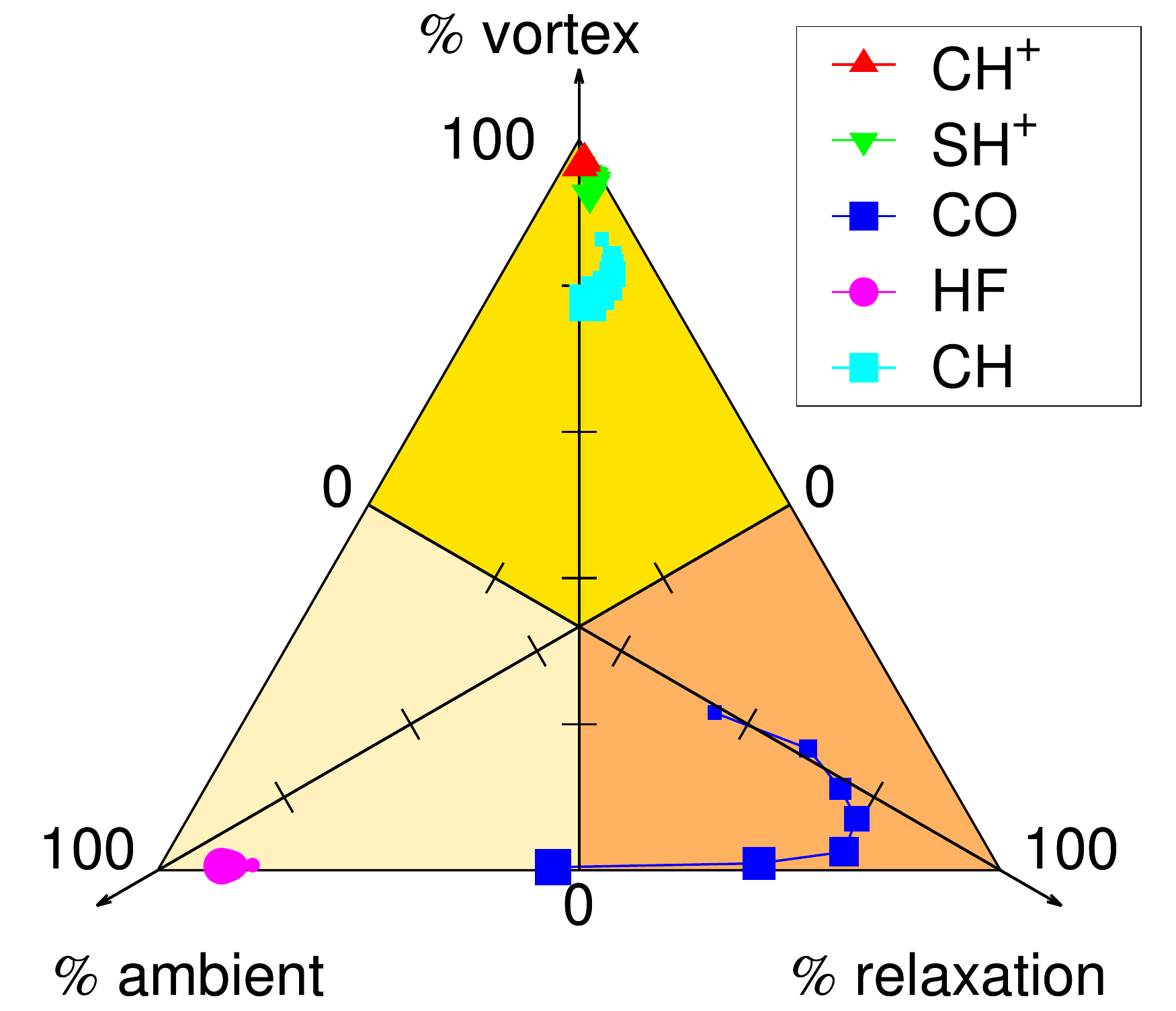}}
\hspace*{1.5cm}
\resizebox{5cm}{!}{%
\includegraphics{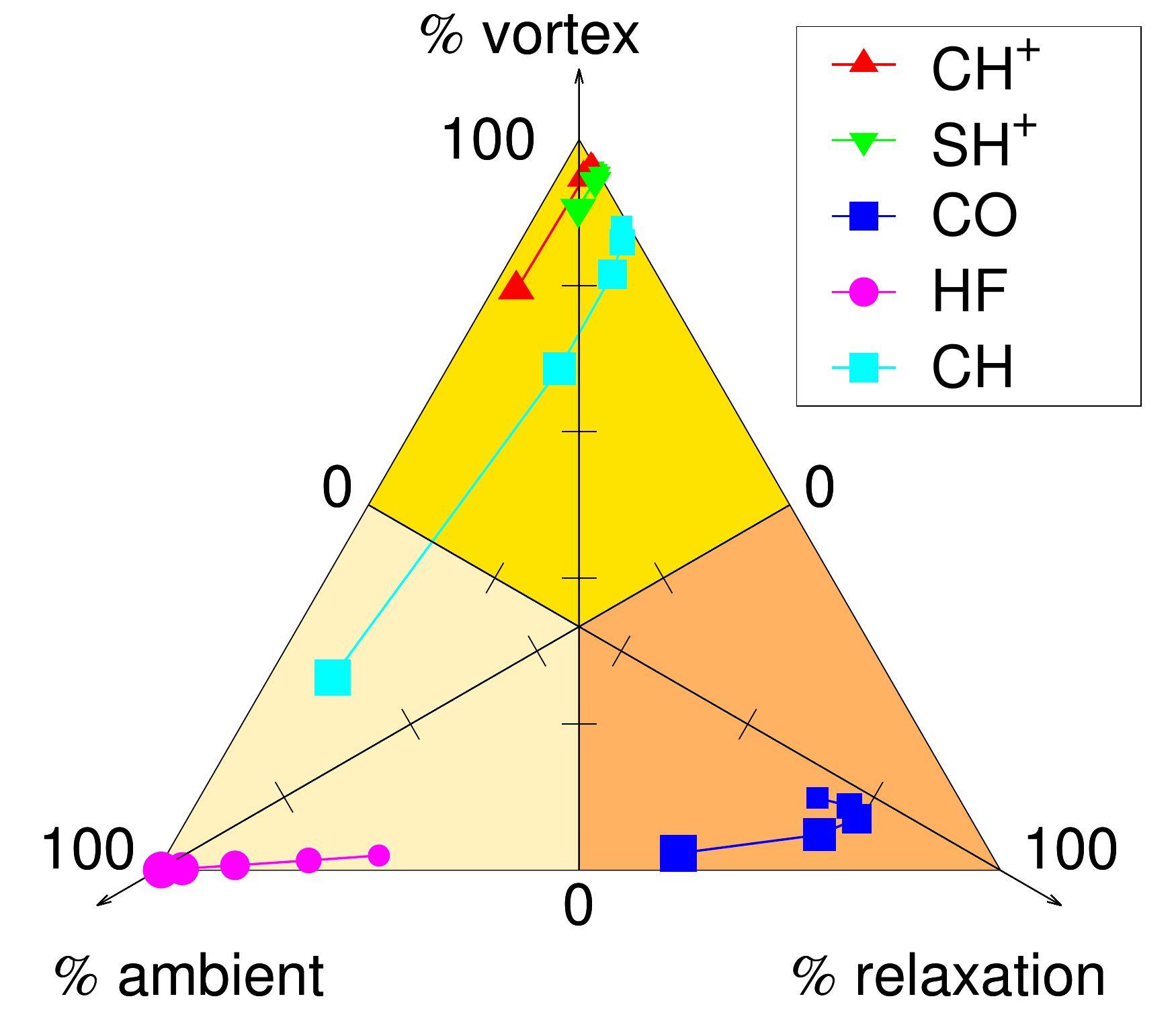}}
\caption{Ternary diagram for the TDR model of \cite{godard:14}, indicating the fraction of each of five diatomic molecules found in three 
components (see text) of a diffuse cloud. The left panel displays model results for a gas density $n_{\rm H} = 50\,\rm cm^{-3}$, and for several values of the visual extinction, $A_{\rm V}$: 0.1, 0.2, 0.3, 0.4, 0.6, 0.8, and 1.0 mag in order of increasing symbol size. The right panel displays results for several values of the density, $n_{\rm H}$: 20, 30, 50, 100 and 300 cm$^{-3}$, at  $A_{\rm V}$ = 0.4 mag.}
\label{fig:G14f3a}
\end{figure}
Turbulent dissipation regions  \citep[TDR;][]{godard:09,godard:12,godard:14} represent an alternative but related scenario in which endothermic reaction rates can be enhanced as a result of locally-elevated gas temperatures and ion-neutral drift. \rev{In this scenario, molecules are produced in turbulent dissipation regions, represented as vortices in which the ion-neutral drift is induced by the ambient magnetic field. Model parameters include the physical conditions in the ambient medium ($n$, $\chi$, A$_V$, $\zeta$) along with four adjustable variables describing the turbulence dissipation rate, the ion-neutral drift, the
vortex magnitude and lifetime. Given reasonable assumptions about the  parameters that describe the turbulence, } TDR models have been successful in accounting for the observed abundances of CH$^+$, SH$^+$ and SH in diffuse molecular clouds.  \textbf{Figure \ref{fig:G14f3a}}, a ternary diagram from the recent study of \cite{godard:14}, shows -- for five diatomic molecules -- the relative fractions of the total column density in three separate components: (1) ambient material currently unaffected by turbulent dissipation; (2) material within the vortices that are the assumed sites of turbulent dissipation in this model \rev{(``vortex'')}; (3) material that has passed through a dissipative vortex \rev{and is currently  relaxing to the ambient chemistry (``relaxation'')}.  This model indicates that the, CH$^+$ and SH$^+$ column densities are dominated by material in dissipative vortices (component 2); the HF column density, by contrast, originates primarily in the ambient gas (component 1).  \rev{The behavior of CH depends  on the density, with the dominant component  switching from vortex (component 2) to the ambient gas (component 1)}.The different behaviors predicted for these four hydrides reflect the fact that the production of CH, CH$^+$ and SH$^+$ is endothermic, while that of HF is exothermic. \rev{While the extensive chemical calculations leading to the results shown in 
Fig. \ref{fig:G14f3a} are based upon a simple characterization of interstellar turbulence with several adjustable parameters, a complementary approach makes use of more sophisticated numerical simulations of turbulence \citep{myers:15}.  To date, studies adopting the latter approach have included a more limited chemistry focused on the CH$^+$ problem}.

\begin{marginnote}[
]
\entry{TDR}{Turbulent Dissipation Region}
\end{marginnote}

In both shocks and TDRs, the dynamical timescales can be short compared to the time needed to establish a chemical equilibrium in which the rates of molecular formation and destruction are equal; a time-dependent treatment of the chemical rate equations is therefore needed.  Simple shock models are sometimes described as "steady-state" models if the physical and chemical conditions are non-varying in a particular inertial frame, but the meaning here is different from "chemical quasi-equilibrium" in which the net rates of formation and destruction are nearly equal for each species.   We note also in this context that even "chemical quasi-equilibrium", which is often a good approximation where shocks and TDRs are absent, is quite different from thermochemical equilibrium (TE), in which the rate of every chemical process equals that of its {\it reverse process} and the molecular abundances are determined solely by thermodynamics.  The condition of TE {\it never} applies in the ISM, because of the presence of cosmic-rays and -- in diffuse and translucent interstellar clouds -- of UV radiation that would be entirely absent in thermal equilibrium at the gas kinetic temperature.

\section{\label{sec:dense}HYDRIDES IN DENSE AND SHIELDED GAS}

\begin{figure}
\includegraphics[width=10cm]{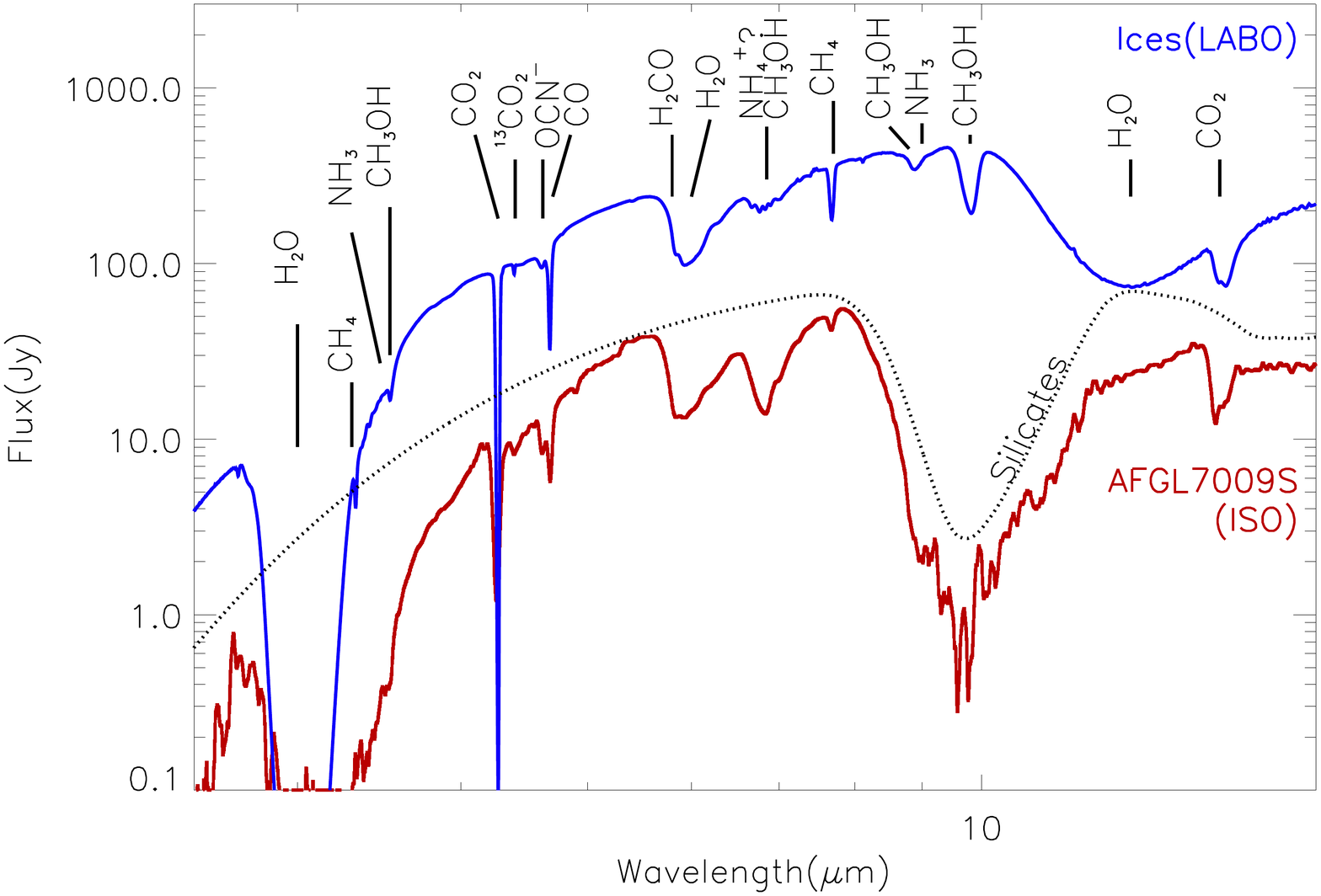}
\caption{\label{fig:ices}Comparison between the {\sl ISO-SWS} spectrum toward the massive protostar RAFGL7009S (red line)
 and a laboratory transmission spectrum  (blue line). The dotted lines shows 
the dust  continuum including absorption from silicates. 
The positions of the main ice constituent absorption lines are indicated \citep[adapted from][]{dartois:98}.}
\end{figure}
Many neutral hydrides stay abundant
in shielded regions, even in dense prestellar and protostellar
cores.  At the low gas and dust temperatures of these regions ($\sim 10$~K), molecules 
freeze on dust grains, building an ice layer. Ice mantles also grow thanks to the condensation of atoms, 
which, by hydrogenation, contribute to the formation of saturated hydrides: H$_2$O for 
oxygen, CH$_4$ for carbon and NH$_3$ for nitrogen. \textbf{Figure \ref{fig:ices}} shows a 
representative IR spectrum of a massive star forming region with prominent absorption features tracing the composition of 
the ice mantle. Water ice is the dominant species, with NH$_3$ and CH$_4$ reaching a few percent of the water 
ice \rev{\citep[][]{boogert:15}}.
 
The low temperatures ($T \sim 10$~K) and absence of FUV radiation represent a favorable 
environment  for fractionation, the chemical process leading to a higher
fraction of an isotope in a given molecule (an isotopologue) than the  ratio of the elemental abundances. The most  efficient 
process is deuterium fractionation, in which one or several hydrogen atoms are 
substituted by deuterium atoms.  As reviewed by \citet{caselli:12b}, deuterium fractionation can be very efficient in 
cold regions, with several molecules reaching abundance ratios [XH]/[XD] exceeding one percent, while the elemental 
D/H ratio is about $2\times 10^{-5}$.  Fractionation may also involve
other isotopes like $^{13}$C, $^{18}$O, $^{17}$O or $^{15}$N.  
Fractionation in D or $^{15}$N are studied to relate ISM processes
with the formation of the first solids in the solar system, since 
primitive material shows localized D and $^{15}$N enhancements \citep{aleon:10}.

In the dense gas around protostars ($n({\rm H_2}) > 10^7$cm$^{-3}$), the dust temperature increases up
to the ice sublimation threshold \rev{($T \sim 100$~K for water ice)}. The gas phase abundances of neutral hydrides
 significantly  increase in these hot cores (for massive stars) and
hot corinos (for low mass stars). \rev{These objects exhibit a  rich spectrum, including numerous hydride emission lines at  (sub)millimeter wavelengths  (e.g. Figure \ref{fig:pacs}), and absorption lines in the infrared, revealing a high abundance of water vapor \citep{boonman:03}, methane and CH$_3$   \citep[\eg][]{knez:09} in the hot gas within a few hundred AU of the protostar. Highly excited NH$_3$ inversion lines at cm wavelengths  also trace the same region \citep{goddi:15}. Such spectra provide   key information about the temperature and velocity field in the vicinity of the protostars.  }

\rev{Hydrides, and especially water and OH, are interesting probes of protoplanetary disks, since different lines can be used to probe different regions : ground state lines trace the cold outer regions of the disk, while excited lines trace the illuminated  surface closer to the central protostar. Following the first detections of water vapor in the inner region (within a few AU) \citep[e.g.][]{salyk:08,carr:08}, {\it Herschel}/PACS spectroscopy in the FIR has confirmed the presence of  OH and H$_2$O in protoplanetary disks around both TTauri and Herbig AeBe stars, the latter having a higher OH/H$_2$O abundance ratio than the former \citep{fedele:13}. Both ortho and para 
H$_2$O ground state lines have been detected by {\sl Herschel}/HIFI towards TW-Hya 
 \citep{hogerheijde:11}. Most of the water is frozen as ice in the disk mid-plane, and water vapor represents a very minor fraction of the water reservoir. Because of the high line opacity, the interpretation of the detected signals requires a good understanding of the disk geometry and sophisticated radiative transfer
calculations.  Other {\sl Herschel} detections include CH$^+$ tracing the UV-irradiated disk surface, and $o$-NH$_3$ associated with cool layers  \citep{thi:11}.}

\begin{marginnote}[
]
\entry{JWST}{The James Webb Space Telescope, a 6.5\,m telescope, operating between 0.6 and 28 $\mu$m, will be launched  in late 2018.
}
\end{marginnote}

\subsection{Water and oxygen hydrides}

\rev{In dense gas, water is the main oxygen hydride, together with hydroxyl radical OH, and their deuterated counterparts HDO, D$_2$O, OD, while the main ion is H$_3$O$^+$. 
While water ice is ubiquitous in molecular \rev{clouds}, with a detection 
threshold  of about 3
magnitudes of visual extinction \citep{whittet:13}, the detection of water vapor in cold cores is very challenging because of the need of a space-borne 
system combining  high  sensitivity and high spectral resolution. Thanks to {\it Herschel}/HIFI, water vapor was detected towards
the L1544 prestellar core \citep{caselli:12}. The line profile shows an inverse P-Cygni profile, with a  blue-shifted emission peak and a red-shifted absorption tracing the slow gravitational collapse of this core. Although
the  water abundance relative to H$_2$ is very small, $\sim 10^{-9}$, the mere presence  of 
water vapor in such a cold ($T =7$~K) and dense ($n({\rm H_2}) > 10^6$~cm$^{-3}$) environment 
indicates that the desorption  induced by the secondary FUV photons \citep[$\chi = 10^{-4}$,][]{prasad:83} produced  during the 
interaction of cosmic rays with molecular hydrogen is efficient. This observation is consistent with predictions from 
chemical models such as those shown in Figure \ref{fig_pdr_mod_hol}, where water vapor is  present at a cloud depth of a few
magnitudes of extinction \citep{hollenbach:12}.
In dense and cold gas, water is cycled between the gas and solid phase  through H$_3$O$^+$. Hence the water OPR is expected to
deviate from thermal equilibrium in a similar way as for NH$_3$ \citep{sipila:15}. Observations of H$_3$O$^+$ or of
its deuterated counterparts  H$_2$DO$^+$, HD$_2$O$^+$ and D$_3$O$^+$ could provide further constraints on the
  cosmic ray ionization rate.
While no detection towards cold cores has been achieved so far,  several H$_3$O$^+$ lines show  up in emission
towards massive star forming regions tracing the presence of 
energetic radiation and/or cosmic rays produced by the embedded protostar \citep{benz:13}. }

\rev{
Although most  rotational water lines are not observable from the ground, 
because of the presence of water vapor in the Earth atmosphere, the transmission is good for rotational lines of heavy water. Both
 HDO \citep{beckman:83} and D$_2$O \citep{butner:07} are present with a moderate level of fractionation. 
 The number of measurements of HDO/H$_2$O and D$_2$O/HDO abundance ratios in the same source is small. In Orion-KL, the prototypical hot-core, HDO/H$_2$O $\sim 3 \times 10^{-3}$, a factor of two larger than D$_2$O/HDO \citep{neill:13}. Most species show a D/H ratio in the range  $2 - 8 \times 10^{-3}$ in this source. As reviewed by \cite{caselli:12b}, the deuterium fractionation of water vapor  is a complex phenomenon, because it is controlled by both solid-phase and gas-phase processes and is likely time-dependent given the similarity of the chemical time scales with the free fall time of dense cores.
The OD radical has been detected with the  SOFIA airborne observatory in the cold envelope surrounding the low mass protostar IRAS16293-2422 \citep{parise:12}. The large OD/HDO abundance ratio (OD/HDO = 20 - 90) 
remains to be fully explained. The  dissociative recombination of H$_2$DO$^+$ may contribute to the OD enhancement, 
but a quantitative assessment awaits a measurement of the branching ratios toward HDO and OD.
Because of their large critical densities and upper energy levels, the sole HDO 
transitions detectable in cold cores are the two  ground state transitions at 465~GHz ($1_{01}-0_{00}$) and 894~GHz ($1_{11} - 0_{00}$).
In  hot cores and hot corinos the conditions are favorable for collisional excitation of 
many HDO levels, leading to a richer spectrum.
For a  collapsing core around a massive protostar, the HDO  ground state transitions are particularly sensitive 
to the velocity field  in the collapsing envelope while the excited lines probe the hot inner region \citep[e.g.][]{coutens:14a}.
Like H$_2$O, D$_2$O has two spin symmetry states, with
 $p$-D$_2$O being the more stable species.  $o$-D$_2$O has its lowest rotational state at a energy  17~K above that of $p$-D$_2$O,  and ortho-states have a  spin degeneracy twice that of para-states. From {\sl Herschel} observations  of both $p$-D$_2$O and $o$-D$_2$O, \cite{vastel:10} concluded that the OPR is consistent with the statistical value of 2.}

\subsection{Methane and carbon hydrides}
\rev{As it is for oxygen hydrides, the competition between condensation and desorption is
 also important for carbon hydrides in dense and FUV-shielded gas. There are, however, two important differences  with oxygen : 
{\it i)} the carbon reservoir is CO and not methane, {\it  ii)} carbon hydride radicals (CH, CH$_2$, CH$_3$) and ions (CH$_3^+$)  initiate the organic chemistry and therefore play a pivotal role
in the formation of a large fraction of the  molecules present in the ISM.  
Another interesting property of the  methylidyne radical, CH, is the linear scaling of its column density,
 deduced from the intensity of its $\Lambda$-doubling transitions at 3.3~GHz,
  with that of
 H$_2$  up to A$_V$ $\sim 3$~mag \citep{mattila:86,chastain:10}. Although these lines  are inverted, with negative excitation 
temperatures,   their integrated intensity approximately 
scales with the CH column density \citep{liszt:02}.  {\it Herschel} observations of
the CH ground state rotational lines confirm this behavior: towards the low mass protostar IRAS16293-2422, 
CH probes the cold envelope \citep{bottinelli:14}, while in more massive sources CH shows a combination of emission 
and absorption features \citep{gerin:10}, indicating that CH is present in the dense material.
The methylene radical,  CH$_2$, shares the same chemistry as 
CH. Its ground state transition has been detected by ISO \citep{polehampton:05} following
its identification in Orion-KL \citep{hollis:95}. }

\rev{Methane is best probed  by infrared spectroscopy using either ground-based or space-born 
telescopes in the 8~$\mu$m region  \citep{lacy:91,oberg:08}. Frozen methane has an abundance $\sim5\%$ that 
of water ice, supporting a formation process through sequential hydrogenation of carbon on grain surfaces. 
The gas phase abundance is more uncertain and ranges between 10$^{-8}$ and 10$^{-6}$ relative to H$_2$. Even for 
the highest abundances, most of the interstellar methane is in solid form.}

\rev{Deuterium fractionation is predicted to be very efficient for carbon hydrides, since deuterated methylium 
CH$_2$D$^+$ is one of the main actors in enriching molecules in deuterium. CH$_2$D$^+$ is operating in lukewarm
 regions ($T  \sim 30 - 70$~K) where the kinetic temperature is too high for H$_2$D$^+$ to be enriched. The DCN and 
HDCO molecules are  typical products of CH$_2$D$^+$ induced chemistry, with observed fractionation levels consistent 
with chemical models  \citep{roueff:13}. Unfortunately, no astronomical detection of the related species  CD and CHD 
has been reported so far. Further tests involve the search for CH$_2$D$^+$.  Its rotational spectrum has been measured 
by \cite{amano:10a}, and a tentative detection has been reported towards Orion-KL by \cite{roueff:13}, 
which awaits confirmation from observations of a less spectrally crowded region.  Because of its small dipole moment
  of 0.3~Debye, CH$_2$D$^+$ 
rotational lines are  weak at the predicted abundance level.
 Deuterated methane CH$_3$D has a non-zero albeit small dipole moment (0.0057~D), allowing observations of rotational transitions from the ground.  After a very deep integration, \cite{sakai:12b} reported
a tentative detection towards the IRAS04368+2557 protostar in the L1527 core. Further searches for CH$_3$D  are needed to confirm this detection, which implies rather large CH$_3$D ($\sim 3 \times 10^{-7}$) and CH$_4$   ($\sim 4 \times 10^{-6}$) gas phase abundances, and very little methane left on grain mantles. }

\subsection{Ammonia and nitrogen hydrides}
\begin{figure}
\includegraphics[width=12cm]{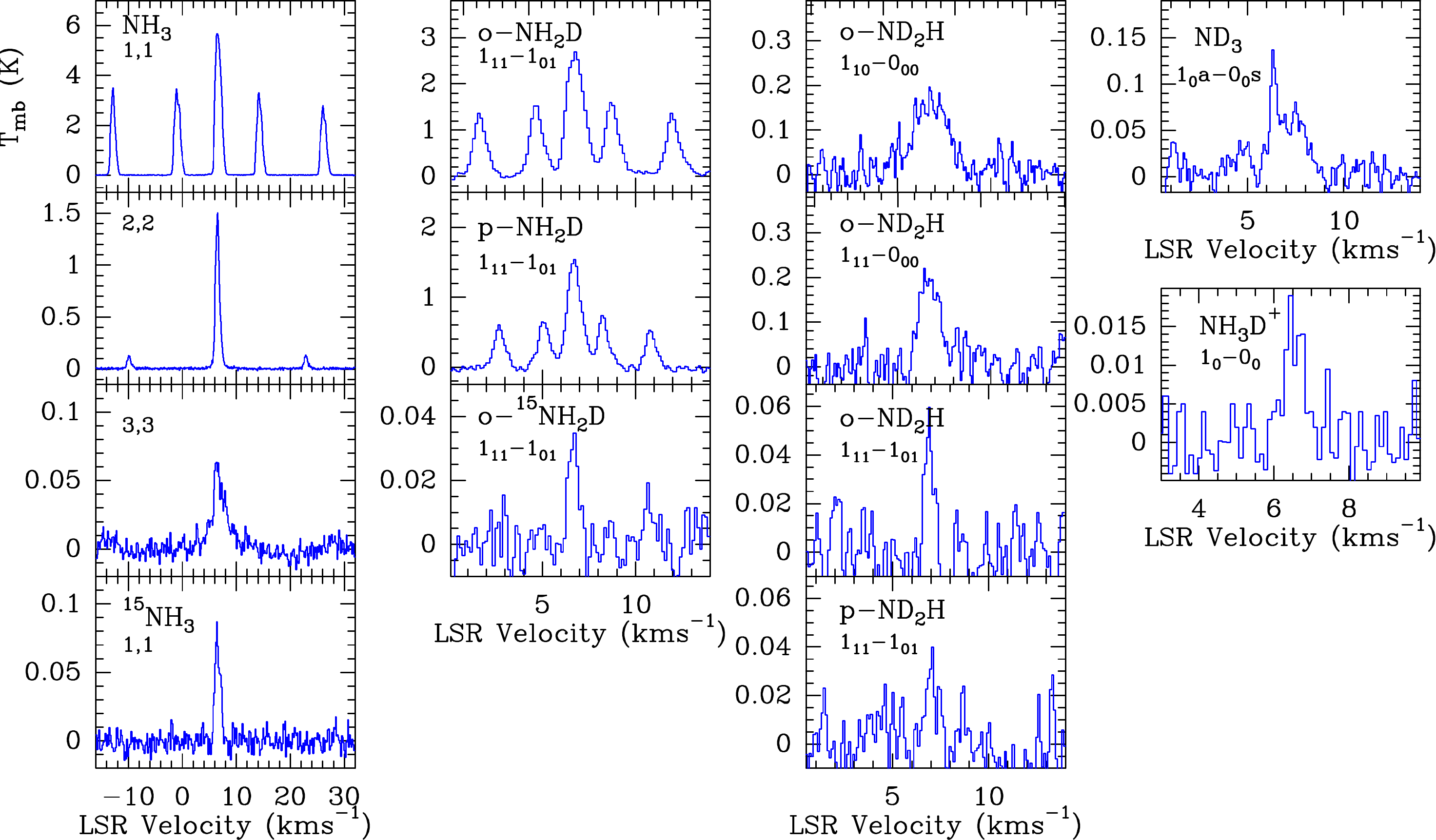}
\caption{Ground based spectra of nitrogen hydrides towards the protostellar core
Barnard1b ($T = 12$~K, $n_{\rm H_2} = 10^5$~cm$^{-3}$, $N(\rm H_2) = 10^{23}$cm$^{-2}$) \citep[adapted from][]{daniel:13,lis:10c,cernicharo:13}.}
\label{fig:nh3}
\end{figure}

Nitrogen hydrides  are mainly formed in the gas phase through a chain of hydrogen abstraction reactions 
starting from N$^+$ + H$_2$ $\rightarrow$ NH$^+$ + H. This reaction
is slightly endothermic (\rev{Fig.~\ref{fig:thermochem}}), and therefore is more efficient for $o$-H$_2$ than for 
$p$-H$_2$. \rev{Because of the rather low formation rate in $p$-H$_2$ enriched gas, a formation route on grain 
 surfaces has also been suggested}.  Subsequent hydrogen abstraction reactions NH$_{n}^+$ + H$_2$ 
$\rightarrow$ NH$_{n+1}^+$ + H with $n = 1,2,3$ followed by dissociative
recombination reactions lead to the molecular ions NH$_2^+$, NH$_3^+$ and NH$_4^+$,
 and the neutral hydrides NH, NH$_2$ and NH$_3$. The detection of NH$_3$D$^+$ 
by \cite{cernicharo:13} nicely confirms this theoretical scheme. 
NH is also formed as a secondary product of  the   dissociative recombination reaction of N$_2$H$^+$  \citep{vigren:12}. 
In dense clouds, the  abundances of nitrogen hydrides are dominated
by ammonia, with NH:NH$_2$:NH$_3$ = 3:1:19 for the dense core around the  protostar IRAS16293-2422 \citep{legal:14}. The excess of NH over NH$_2$ is explained by its 
additional formation route from the dissociative recombination of N$_2$H$^+$. 

Ammonia has two spin symmetry states, $o$-NH$_3$ with $K$ values 
multiple of three and $K=0$, and $p$-NH$_3$ for other $K$ values. 
The energy difference between $o$-NH$_3$ and $p$-NH$_3$ is 21.5~K, and  the high temperature value of the OPR ratio
 is 1, while an excess of $o$-NH$_3$ is expected at temperatures lower than 50~K.  
As a symmetric top with two equilibrium configurations, ammonia 
 exhibits inversion transitions between equivalent rotational states. 
While the low energy NH$_3$ inversion transitions probe the para species, the
first $o$-NH$_3$ state accessible from the ground is the $(J,K) = (3,3)$ level at 
120~K above ground. Therefore most  NH$_3$ measurements  in dense cores
 only trace $p$-NH$_3$, and the total ammonia column density is
derived assuming an ortho-to-para ratio (OPR) of one.  
Using {\sl Herschel} absorption data on $o$-NH$_3$ and $p$-NH$_3$, \citet{persson:10} measured an  anomalous value of the OPR ratio of 0.7.   In $p$-H$_2$ enriched gas, the chain of ion-molecule reactions forming NH$_3$ favors the formation 
of $p$-NH$_3$ due to spin statistics selection rules \citep{rist:13,faure:13}. This effect could be
prominent in cold cores, with predicted  OPR $\sim 0.5$ for NH$_3$. The deficit
in $o$-NH$_3$ implies \rev{a small error} in the total NH$_3$ column density deduced from $p$-NH$_3$ observations.
 The amidogen radical NH$_2$ is formed in the same chain of
gas phase reactions as ammonia. With two identical protons, NH$_2$ has two spin 
symmetry states,  the ortho state with a spin multiplicity
of three is the most stable, and the para state with a spin multiplicity of one, with an energy difference of 30~K. As for ammonia, the NH$_2$ OPR  is sensitive to the presence of an excess of $p$-H$_2$, and can reach values lower than 
3, which are not permitted in thermal equilibrium, in $p$-H$_2$ enriched gas \citep{legal:14}. 
So far only a handful of measurements of the NH$_3$ and  NH$_2$ OPR 
are available \citep{persson:15}. Such measurements may be able to put 
constraints on the H$_2$ OPR in dense and shielded gas.

\rev{  Ammonia is a a good tracer
of both prestellar and protostellar cores, with a fractional
 abundance relative to H$_2$ of $\sim 10^{-8}$  \citep{bergin:07}. It 
remains undepleted in the gas phase up to relatively large 
densities ($\sim 10^6$ cm$^{-3}$), in contrast to the case of CO.  As illustrated in \textbf{Figure \ref{fig:nh3}}, all
deuterated ammonia isotopologues have been detected
in the ISM : NH$_2$D, ND$_2$H and ND$_3$ \citep{roueff:05}, as well as the $^{15}$N variants $^{15}$NH$_3$  \citep[\eg][]{lis:10c} and  $^{15}$NH$_2$D \citep{gerin:09}, and even NH$_3$D$^+$ \citep{cernicharo:13}. 
The deuterated ammonia isotopologues are better
 tracers of the cold  inner regions of prestellar cores, where most molecules 
are frozen onto grains, than is NH$_3$ itself, with fractionation levels  above 10\%  
in such regions \citep[\eg][]{daniel:13}.
The fractionation of ammonia and its precursors NH and NH$_2$ 
is most efficient at temperatures lower than $\sim 20$~K 
because it involves reactions with HD, H$_2$D$^+$, D$_2$H$^+$ and D$_3^+$ that are enhanced at temperatures below 20~K \citep{roueff:15}. ND has been detected  \citep{bacmann:10}, with  [ND]/[NH] reaching $\sim 10$~\%, a value  similar to that 
observed for [NH$_2$D]/[NH$_3$], but with a large uncertainty because of the difference 
in excitation between NH and ND.}

At low temperatures, nitrogen  may  become fractionated, leading to 
a possible difference between the elemental isotopic ratio $^{15}$N/$^{14}$N 
and the abundance ratio of nitrogen-bearing species. The topic is of interest 
because  analysis of primitive solar system material as found in carbonaceous chondrites or in micro-meteorites have led to the identification of 
specific regions with isotopic ratios
very different from the bulk, in either D/H, $^{15}$N/$^{14}$N or both \citep{aleon:10}. 
The difference between the $^{15}$N/$^{14}$N isotopic ratio  in the Sun 
(and the solar system), $2.3 \times 10^{-3}$, and that of the Earth, $3.8 \times 10^{-3}$, also indicates that active fractionation processes were operating in the early solar system. 
Following the evolution of D/H and $^{15}$N/$^{14}$N in interstellar molecules 
along the formation of a protostar and its circumstellar disk is therefore a 
means of constraining the chemical evolution of the star forming system and the formation 
of its primitive bodies.
So far, measuring the $^{15}$N/$^{14}$N isotopic ratio has proven to be much
more difficult than D/H because of the large dynamical range required.
Chemical models have investigated the coupled fractionation of hydrogen,
carbon, and nitrogen, and also the dependence with time during the collapse.
The prediction is that NH$_3$ should preserve within 20\% the original nitrogen isotopic
ratio, unlike other species such as HCN which could become highly fractionated \citep{wirstrom:12,roueff:15}. 
\rev{Selective photodissociation of N$_2$ as modeled by \cite{heays:14} also plays a role in the nitrogen fractionation in protoplanetary disks. }

\subsection{Sulfur and other heavy element hydrides}
 Hydrogen sulfide H$_2$S is the main sulfur hydride in dense clouds, while SH and SH$^+$ are more confined to diffuse and translucent clouds \citep{neufeld:15a,crockett:14}, or the UV-illuminated cloud edges as described in Section \ref{sec:pdr}. The sulfur chemistry proceeds differently from that
of oxygen or carbon because the hydrogen abstraction reactions between HS$_n$$^+$ and H$_2$ are highly 
endothermic. The  hydrogenation
of sulfur atoms on grains, followed by desorption, is a possible chemical pathway to form hydrogen sulfide in dense cores, although no confirmed detection of H$_2$S ice has been reported so far.
Singly and doubly deuterated hydrogen sulfide have been detected in a
few dark clouds \citep{vandishoeck:95b,vastel:03} with a significant 
fractionation level of
 $\sim 10\%$. The fractionation is sensitive to the temperature since
HDS and D$_2$S are not detected in hot cores, despite their large H$_2$S 
column densities. For the Orion-KL region, the upper limit for the HDS/H$_2$S 
ratio is a few 10$^{-3}$, at the
same level as HDO/H$_2$O \citep{crockett:14}.

The halogen hydrides HF and HCl are expected to be the main gas phase reservoirs
of fluorine and chlorine  in  dense and well shielded gas \citep{neufeld:09}, and could be used to monitor
the depletion of these elements.  The detection of the $J=2-1$ line of HF in absorption towards SgrB2 \citep{neufeld:97} implied a fluorine depletion close to two orders of magnitude. Surveys of hydrogen chloride also conclude
 that chlorine depletes on grains by at least two orders of magnitude \citep[\eg][]{peng:10,kama:15}. While 
many previous analysis relied on collisional cross sections with He, new calculations have been performed
 for the  H$_2$-HCl system by \cite{lanza:14}, providing further confirmation of the low
HCl abundance in dense gas. HCl may return to the gas phase in the hot inner region around
massive protostars \citep{goto:13}.

 \section{HYDRIDES AS DIAGNOSTIC TOOLS}
\label{sec:diagnostic}
Because the chemical pathways leading to the formation of interstellar hydrides are relatively simple and well-understood, the analysis of the observed abundances is relatively straightforward and provides key information about the physical and chemical conditions within the environments in which hydrides are found.  Several examples have already been discussed.  These include observations of CH$^+$ and SH$^+$ in diffuse molecular clouds, which trace the effects of shocks or TDRs (see Section \ref{sec:shocks}), and observations of neutral hydrides such as HF, HCl and H$_2$O in dense clouds, which probe the depletion of heavy elements onto icy grain mantles (see Section \ref{sec:dense} and \ref{sec:chemistry}).
In this section, we will discuss several additional cases where observations of hydride molecules provide unique information about the ISM.  In particular, we will consider how observations of specific hydrides -- namely CH, H$_2$O, and HF -- provide valuable surrogate tracers for molecular hydrogen, the dominant molecular constituent of the ISM but one that is difficult to observe directly, while several hydride cations (including ArH$^+$, HCl$^+$, OH$^+$ and H$_2$O$^+$) are most abundant in {\it partially-molecular} gas, with each molecule predicted to show a peak abundance in material with a specific molecular fraction, $f_{\rm H_2}$ 
We will also discuss how hydride molecules can measure the cosmic-ray (or X-ray) ionization rate in the Milky Way and other galaxies. 

\subsection{Hydride as tracers of H$_2$}  

Although it is the primary molecular constituent of the ISM, H$_2$ is hard to detect because it lacks a permanent dipole moment.   Absorption line observations of the ultraviolet (and dipole-allowed) Lyman and Werner bands of H$_2$ \rev{measure the H$_2$ column density directly}, although they are limited to relatively unreddened lines-of-sight toward relatively nearby hot stars \citep[e.g.][]{rachford:09}.  Absorption in the near-infrared quadrupole-allowed vibrational bands has also been observed toward a few highly-embedded infrared sources \citep{lacy:94,kulesa:02}, but is an extremely insensitive probe (in that very large H$_2$ column densities are needed to generate even very small optical depths).
 There has therefore been a considerable effort to identify surrogate observational tracers that can be used to measure H$_2$ column densities and masses.  Observations of CO emission have been widely used to measure H$_2$ masses \citep[][and references therein]{bolatto:13}, but fail to trace a significant component of the molecular ISM.  This component has been called the ``CO-dark" or "CO-faint" molecular gas \citep{grenier:05, wolfire:10, bolatto:13}, and consists of material that is sufficiently shielded from UV radiation to have a substantial H$_2$ fraction but not so well-shielded as to contain abundant CO. \rev{This material lies at the transition between diffuse and translucent 
clouds near $A_V = 1$ mag \citep[e.g.][]{ewine:98}}.

Three hydride molecules -- CH, H$_2$O, and HF -- exhibit remarkably constant abundance ratios relative to H$_2$, and are therefore valuable as surrogate tracers that are sensitive to CO-dark gas.  The best studied case is CH, for which optical absorption-line observations can be calibrated with respect to UV absorption-line observations of H$_2$, yielding $N({\rm CH})$ and  $N({\rm H}_2)$ column densities for sight-lines with $N({\rm H}_2)$ in the range $\sim 10^{19}$ to $ \sim 10^{21}\rm cm^{-2}$ \citep[e.g.][%
]{sheffer:08}. The observed $N({\rm CH})/N({\rm H}_2)$ ratio  has a mean of $3.5 \times 10^{-8}$, shows a dispersion of 0.21 dex, and exhibits no dependence on $N({\rm H}_2)$; moreover, the $N({\rm CH})/N({\rm H}_2)$ can be determined for sight-lines of even larger $N({\rm H}_2)$ -- albeit less directly through a comparison of cm-wavelength CH emission with reddening measurements -- and is found to remain entirely consistent with the $N({\rm H}_2) \simlt 10^{21}\rm cm^{-2}$ value for ${\rm H}_2$ column densities up to $\sim 4 \times 10^{21}\rm cm^{-2}$ \citep{mattila:86,liszt:02}. \rev{State of the art steady-state chemical models 
reproduce the observed trend between CH and H$_2$ correctly for densities larger than 100~cm$^{-3}$, while for lower densities the CH production is closely related to that of CH$^+$, as explained  in Sect~\ref{sec:shocks}. }

Observations of submillimeter absorption, performed with the HIFI instrument on {\it Herschel} have extended the use of hydrides as probes of $N({\rm H}_2)$, providing access to the $N=1, J=3/2 \rightarrow 1/2$ CH $\Lambda$-doublet at 532/536 GHz along with rotational transitions out of the ground-state of the neutral hydrides HF and H$_2$O.  Such observations show the $N({\rm HF})/N({\rm CH})$ ratio to be remarkably constant in diffuse clouds, with a typical value of 0.4, implying a mean $N({\rm HF})/N({\rm H_2})$ ratio of $1.4 \times 10^{-8}$, given a $N({\rm CH})/N({\rm H}_2)$ ratio of $3.5 \times 10^{-8}$.  In the limit where $f_{\rm H_2} \sim 1$, this $N({\rm HF})/N({\rm H_2})$ ratio implies that HF accounts for $\sim 40 \%$ of the gas-phase fluorine. A direct measurement of the $N({\rm HF})/N({\rm H}_2)$ column density ratio in the diffuse ISM has been obtained \citep{indriolo:13} toward a single source, the hot star HD 154368, by combining observations of the near-IR $v=1-0\,\,R(0)$ HF {\it vibrational} transition with those of dipole-allowed UV transitions of H$_2$ .  This measurement provided an independent and entirely consistent ($1.15 \pm 0.41 \times 10^{-8}$) determination of the $N({\rm HF})/N({\rm H_2})$ ratio.  Originally, the predictions of astrochemical models by \cite{neufeld:09} for fluorine-bearing interstellar molecules predicted a $N({\rm HF})/N({\rm H_2})$ ratio roughly twice as large as the observed value.  However, recent laboratory measurements \citep{tizniti:14} of the rate coefficient for the reaction of H$_2$ with F to form HF appear to have revealed the source of the discrepancy: the original astrochemical models adopted a theoretical value for the rate coefficient at low temperature that was a significant overestimate.  Revised models that assume instead the laboratory-measured value for the rate of HF formation predict $N({\rm HF})/N({\rm H_2})$ ratios in good agreement with the astronomical observations.  

\begin{figure}
\hskip -0.1cm
\resizebox{5.cm}{!}{\includegraphics{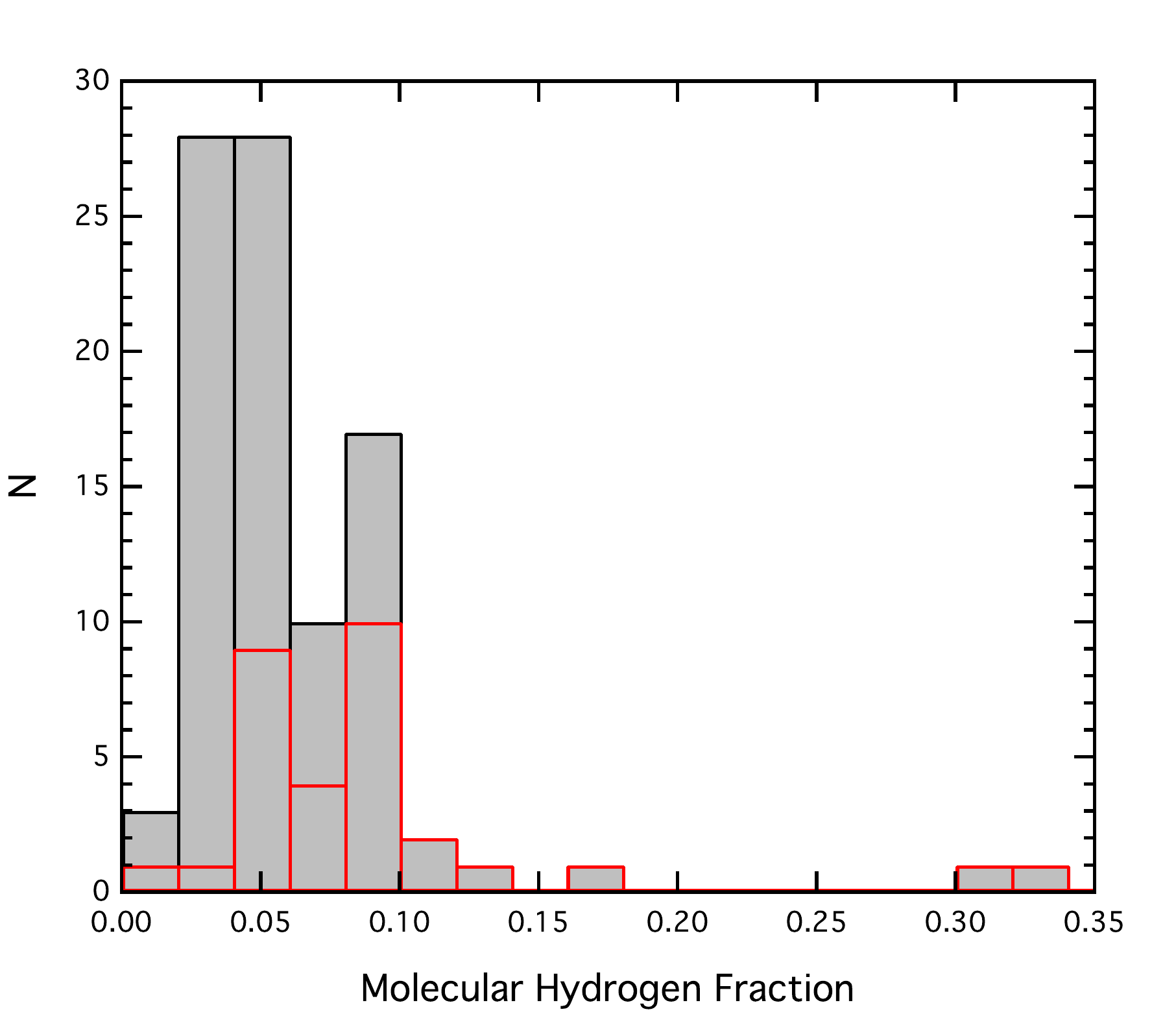}}
\hskip 1.5cm
\resizebox{5.cm}{!}{\includegraphics{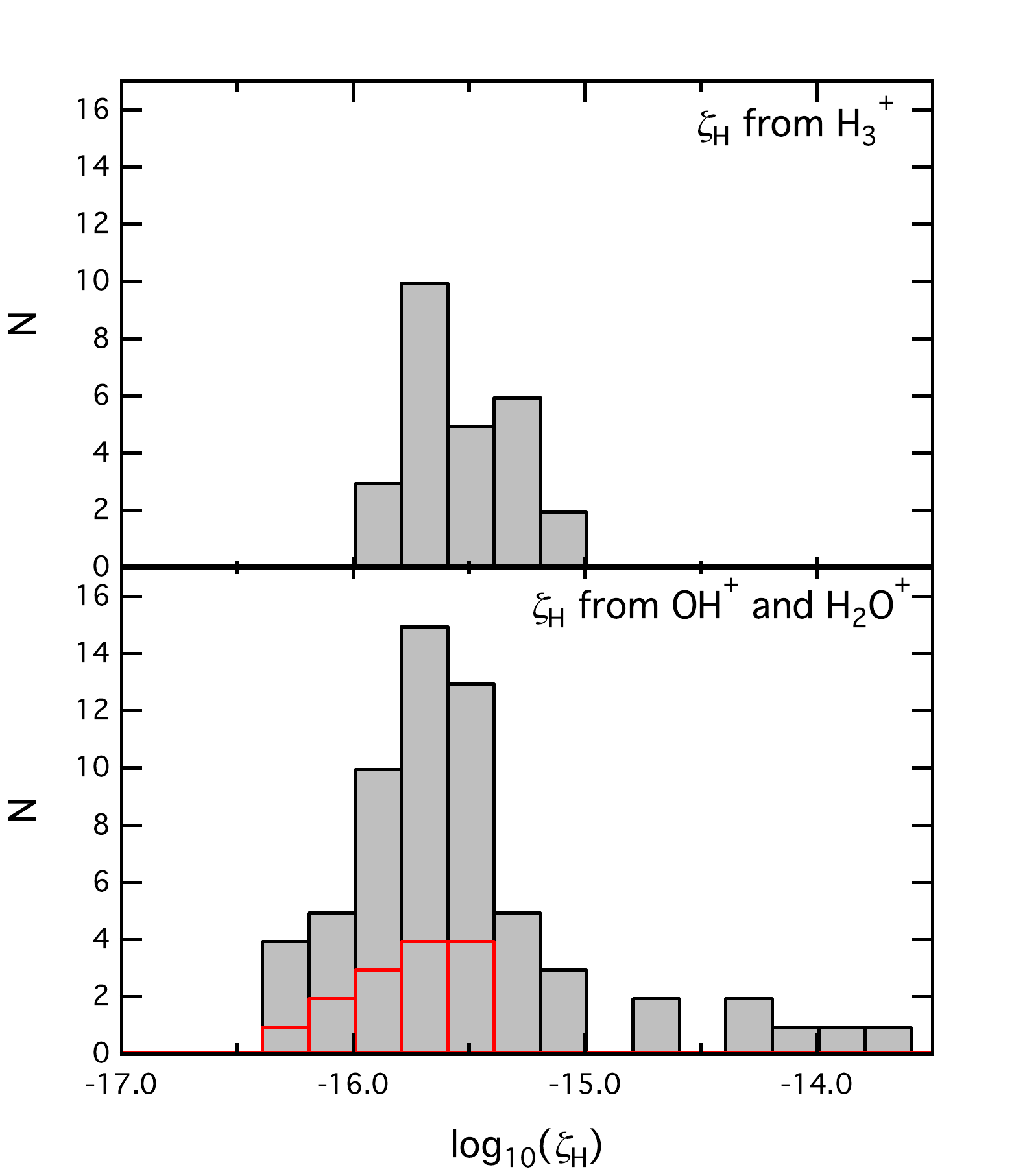}}
\caption{Distributions of molecular hydrogen fraction, $f_{\rm H_2}$ (left panel) and cosmic-ray ionization rate, $\zeta_{\rm H}$ (lower right panel), inferred from {\it Herschel} observations of OH$^+$ and $\rm H_2O^+$ and reproduced from \citet{indriolo:15}.  The red histogram applies to material with line-of-sight velocities within 5~km~s$^{-1}$ of the systemic velocity of the background source (i.e.\ to material that may be in the vicinity the continuum source).  The upper right panel shows the distribution of cosmic-ray ionization rates inferred (for a different set of sight-lines) from observations of H$_3^+$.}  
\label{fig:ohplus_hist}
\end{figure}

Another neutral hydride that has been extensively observed with {\it Herschel}, H$_2$O, is a third potential H$_2$ tracer within the diffuse ISM.  The mean $N({\rm H_2O})/N({\rm HF})$ ratio determined from {\it Herschel} observations of the diffuse ISM is 1.73 \citep{sonnentrucker:15}, and the standard deviation is 0.87, implying a typical $N({\rm H_2O})/N({\rm H_2})$ ratio of $2.4 \times 10^{-8}$.  The variation in the H$_2$O abundance is apparently somewhat larger than those in the CH and HF abundances.  

Taken together, submillimeter absorption line observations of CH, H$_2$O and HF provide a valuable toolset for the determination of $N({\rm H_2})$.  Their column densities lie in the average ratio $N({\rm CH})$ : $N({\rm H_2O})$ : $N({\rm HF})=$ 2.5 : 1.7: 1.0.  For the spectral lines that have been most extensively observed in absorption -- the 536 GHz $N=1, J=3/2 \rightarrow 1/2$ transition of CH, the 1113 GHz $1_{11}-0_{00}$ transition of $p$-${\rm H_2O}$, and the 1232 GHz $J=1-0$ transition of HF -- the corresponding ratio of velocity-integrated optical depths is 0.26:0.6:1.0.  Thus, the HF transition provides the most sensitive probe of H$_2$ along any sight-line, followed by H$_2$O and CH.  These probes are complementary because in cases where the HF absorption is optically-thick, yielding only a lower limit on the column density, the CH absorption is often optically-thin.    
Conversely, absorption features that are too weak to detect in observations of CH may be detectable in observations of HF.  While CH, HF and H$_2$O, can be regarded as ``primary" tracers of molecular hydrogen, their column densities are well correlated with other, non-hydride molecules that can be detected from the ground at millimeter wavelengths, including HCO$^+$ \citep[\eg][]{lucas:96} and C$_2$H \citep{lucas:00,gerin:10}.  Absorption line observations of these non-hydride ``secondary" tracers can also be used to estimate $N({\rm H}_2)$, albeit with lower accuracy. \rev{OH can be added to the list of the ``secondary'' tracers, given the excellent correlation between the opacities of the
 HCO$^+$ $J=1 \rightarrow 0$ and OH 1665 \& 1667 MHz {\it absorption} lines \citep{liszt:96}, and the moderate scatter in the 
OH/H$_2$O column density ratio \citep{wiesemeyer:15}. }

Along with absorption line observations of CH, HF, and H$_2$O, observations of \rev{CH and} OH {\it emission} show considerable promise as a probe of CO-dark molecular gas \citep{liszt:96}.  This promise has been demonstrated recently by \cite{allen:15}, who performed a sparse, high-sensitivity pilot survey near the Galactic plane for the 18 cm emission lines of OH: fewer than one-half of the multiple spectral features thereby detected were accompanied by detectable CO line emission. Further investigations, both observational and theoretical, will be needed to make quantitative use of this H$_2$ tracer.

\subsection{Probes of the molecular fraction}

While the neutral hydrides CH, HF and H$_2$O are most abundant in gas with a large molecular fraction, $f_{\rm H_2}$, several molecular ions show peak abundances in material with a smaller H$_2$ fraction.  This behavior is particularly notable for ions that are destroyed rapidly by reaction with H$_2$: these include ArH$^+$, which is destroyed by proton transfer to H$_2$ \citep{schilke:14}, and OH$^+$ and $\rm H_2O^+$, which undergo H-atom abstraction reactions to form $\rm H_2O^+$ and $\rm H_3O^+$ respectively.  The $n({\rm OH}^+)/n({\rm H_2O}^+)$ ratio is a particularly valuable probe of the H$_2$ fraction \citep{neufeld:10}, because the reaction of H$_2$ with OH$^+$, which forms H$_2$O$^+$, competes with dissociative recombination of OH$^+$ with electrons; thus, the $n({\rm OH}^+)/n({\rm H_2O}^+)$ ratio is larger in regions of small $f_{\rm H_2}$ and smaller in regions of large $f_{\rm H_2}$.  Through an analysis of 92 OH$^+$ and $\rm H_2O^+$ absorption features detected by {\it Herschel} along the sight-lines to 20 bright Galactic continuum sources, \cite{indriolo:15} obtained the histogram of $f_{\rm H_2}$ values reproduced in \textbf{Figure \ref{fig:ohplus_hist} (left panel)}.  This analysis clearly reveals a substantial amount of interstellar gas with a molecular hydrogen fraction of a few percent.  Of course, because the observations measure a ratio of column densities rather than volume density (i.e.\ $N({\rm OH}^+)/N({\rm H_2O}^+)$ rather than $n({\rm OH}^+)/n({\rm H_2O}^+)$), the $f_{\rm H_2}$ values thereby derived are really averages  over the material {\it that contributes most to the OH$^+$ and $\rm H_2O^+$ absorption.}  These $f_{\rm H_2}$ values may therefore differ substantially from those derived for the entire sight-line from a comparison of $N({\rm H_2})$ and $N({\rm H})$.

The argonium ion, ArH$^+$, traces material of even smaller $f_{\rm H_2}$.  Because the dissociative recombination of ArH$^+$ is unusually slow, reactions with H$_2$ become the dominant destruction mechanism for $f_{\rm H_2} \simgt 10^{-4}$ and the ArH$^+$ abundance starts to drop once $f_{\rm H_2}$ increases beyond that value.  Thus, as discussed by \cite{schilke:14}, ArH$^+$ serves as a unique probe of material that is {\it almost purely atomic}.  This prediction for the behavior of ArH$^+$ is supported by the observational finding that the distribution of ArH$^+$ is entirely different from that of the other interstellar hydrides (see Figures \ref{fig:absorption_spectra} and \ref{fig:pca} in Section \ref{sec:diffuse} above).     

\subsection{Probes of the cosmic-ray ionization rate}

Unlike the formation of CH$^+$ and SH$^+$, which is driven by shocks or turbulent dissipation or enhanced in strongly FUV-irradiated gas, the formation of ArH$^+$, OH$^+$ and H$_2$O$^+$ is  initiated by cosmic-rays.  ArH$^+$ formation is driven by the cosmic-ray ionization of atomic argon, with then reacts with H$_2$ to form ArH$^+$; in the case of OH$^+$ and H$_2$O$^+$ formation, the cosmic-ray ionization of H (or H$_2$, in material of large $f_{\rm H_2}$) is the process that initiates the relevant reaction chain.  Observations of the abundances of OH$^+$ and H$_2$O$^+$ in diffuse clouds, supported by detailed astrochemical calculations \citep{hollenbach:12}, have been used to measure the cosmic-ray ionization rate in the Milky Way galaxy \citep{gerin:10a,neufeld:10,indriolo:15}.  

\cite{indriolo:15} performed an analysis of all the relevant data available from {\it Herschel} to obtain the histogram reproduced in Figure \ref{fig:ohplus_hist} (lower right panel), which shows the distribution of cosmic-ray ionization rates in the disk of the Milky Way, $\zeta_{\rm H}$, defined here as the total rate of ionization per H atom.  This quantity shows a log-normal distribution, with log$_{10}\zeta_{\rm H}$ having a mean of --15.75 (corresponding to $\zeta_{\rm H} = 1.8 \times 10^{-16}\, \rm s^{-1}$) and standard deviation 0.29.
The results for $\zeta_{\rm H}$ obtained from OH$^+$ and H$_2$O$^+$ are considerably larger than the canonical values adopted for diffuse clouds by most authors prior to 2002, but are in good agreement with estimates \citep[\eg][]{indriolo:12b} derived (for different sight-lines) from mid-IR observations of H$_3^+$.   The upper right panel in Figure \ref{fig:ohplus_hist} shows the distribution of $\zeta_{\rm H}$ derived from recent H$_3^+$ observations.
 
The ionization rates inferred for material in the Galactic Center are typically 10 -- 100 times as large as the average value inferred for gas in the disk, and even larger ionization rates (exceeding $10^{-13}\, \rm s^{-1}$) have been derived for circumnuclear regions within the ULIRGs Arp 220 and NGC 4418 \citep{gonzalez:13}.  Because ionization by hard X-rays has similar effects on the chemistry as those of cosmic-ray ionization, it is not possible to unambiguously discriminate between these two possible ionization sources based on observations of hydrides alone.    

\subsection{Other diagnostic uses}
 
In addition to those diagnostics that may be exploited by an analysis of hydride {\it chemistry}, observations of hydrides offer other tools that measure key parameters of the interstellar environment.  The oblate symmetric top molecules NH$_3$ and H$_3$O$^+$ can provide physical probes of the gas temperature.   For such molecules, the lowest states of a given quantum number $K$ are metastable (and have a total angular momentum quantum number $J=K$).  These metastable states do not undergo radiative decay, and thus their relative populations are determined by non-radiative processes.  For low-lying metastable states of NH$_3$ and $\rm H_3O^+$, the relative populations are determined by collisional excitation and probe the kinetic temperature of the gas, providing a ``molecular cloud thermometer" \citep[e.g.][]{walmsley:83}.  In the case of high-lying metastable states, however, the effects of ``formation pumping" may complicate the interpretation of the observations \citep[e.g.][]{lis:14}.

The Zeeman splitting of hydride spectral lines offers another important diagnostic tool that has been used to estimate the interstellar magnetic field strength.  The paramagnetic OH radical is of particular value, and absorption line observations of its 18 cm transitions have been used to determine the magnetic field in interstellar gas with densities in the range $n_{\rm H} \sim 10^2 \, \rm{cm}^{-3}$ to $\sim 10^4\, \rm{cm}^{-3}$ \citep[e.g.][for a review]{crutcher:12}.   This material is intermediate between the lower density gas probed by H{\sc{i}} Zeeman measurements and the higher density gas probed by CN Zeeman measurements.  At yet higher densities, the Zeeman splitting of OH and H$_2$O {\it maser} transitions has been used to measure magnetic field strengths, in extreme cases revealing fields in excess of 100~mG.

\rev{Because the 22 GHz water maser emissions from shocked regions can be extraordinarily bright, they may be observed by means of Very Long Baseline Interferometry (VLBI), providing sub-milliarcsecond resolution and the possibility of measuring parallaxes and proper motions with accuracies of $\sim 10\, \mu \rm as$ and $\sim 100\, \mu \rm as\,yr^{-1}$ respectively \citep[\eg][]{reid:09,honma:12}.  Such measurements provide uniquely valuable probes of fundamental Galactic parameters such as the distance to the Galactic Center and the circular rotation speed at the solar circle. }

Finally, observations of hydride isotopologues can be used to determine elemental isotope ratios, which -- in turn -- can provide constraints upon the star formation history and the chemical evolution.  In the diffuse ISM, for example, where fractionation is generally believed to be unimportant for isotopes of elements other than hydrogen or carbon, observations of H$_2^{35}$Cl$^+$ and H$_2^{37}$Cl$^+$ have been used to determine the $^{35}$Cl/$^{37}$Cl isotopic ratio, both in the Galactic disk \citep[and references therein]{neufeld:15b} and in the redshift 0.89 absorber in front of the lensed blazar PKS 1830-211 \citep[][see Section \ref{sec:extragalactic} below]{muller:14b}. All these determinations are entirely consistent with the solar system $^{35}$Cl/$^{37}$Cl isotopic ratio of 3.1 \citep{lodders:03}.

\section{HYDRIDES IN EXTERNAL GALAXIES}
\label{sec:extragalactic}

As relatively abundant and ubiquitous species, hydrides are
detected in external galaxies, from local objects up to high redshift systems.
The list of species includes water vapor \citep{yang:13,omont:13}, OH, 
and the related ions H$_3$O$^+$, 
H$_2$O$^+$ and OH$^+$; NH$_3$, CH and CH$^+$, as well as HF and the chlorine
bearing species HCl and H$_2$Cl$^+$ \citep{vdwerf:10,rangwala:11,rangwala:14, gonzalez:12,gonzalez:13,kamenetsky:12,muller:14b,muller:14a}.
The hydride diagnostic capabilities can be exploited to probe the physical
conditions, turbulence and radiation field including X-rays and cosmic rays in
a more extreme range than what is found in the Milky Way.

\subsection{Local galaxies}
FIR spectroscopy, performed first with {\sl ISO}, then with
 {\sl Herschel} revealed a diversity of spectral lines,  
especially from
hydrides that accompany the set of CO rotational lines and of atomic fine structure lines.
In local galaxies ($z < 0.1$), the strongest rotational lines of water vapor are  the 
$2_{02}-1_{11}$ and $3_{21}-3_{12}$ transitions with 
rest frequencies 987.9 and 1162.9 GHz. Their luminosities are linearly correlated with the far infrared 
luminosity, $L_{FIR}$,  across a broad range of infrared luminosities, 
$1 - 300 \times 10^{10}$ L$_\odot$, i.e. from normal galaxies up to very 
active starburst \citep{yang:13}. Radiative transfer models show that FIR pumping is an important source of 
excitation for most H$_2$O lines, except for
those connected to the ortho or para ground states.
 Fully resolved line profiles have been obtained with HIFI, 
showing the  complexity of the line profiles even in nearby galaxies :
towards M82, different H$_2$O lines have different line profiles ranging from pure absorption to 
pure emission \citep{weiss:10}.

 At moderate spectral resolution ($\sim 200$~kms$^{-1}$), submm and FIR
hydride lines appear in emission, in absorption or even with P-Cygni profiles. In galaxies hosting a compact starburst or
an active galactic nucleus (AGN) the FIR radiation from the hot dust is pumping the rotational
levels, leading to a variety of line profiles revealing the conditions
close to the heavily buried compact source \citep{vdwerf:10}.
 OH lines from excited levels are especially interesting because they trace powerful outflows of molecular 
material \citep{sturm:11,gonzalez:12}, that represent an important source
of negative feedback in starburst galaxies or  AGNs.  In an analysis of 29 galaxies, 
\cite{gonzalez:15} clearly show that the OH 65$\mu$m equivalent width probes the fraction of
 infrared luminosity produced by a compact and warm source, i.e. is closely related 
 to the presence of  an heavily buried active region, which could escape detection otherwise.
 Absorption from 
excited levels of other hydrides like water vapor, OH$^+$ and H$_2$O$^+$ can
also be detected. H$_3$O$^+$ deserves a special treatment. 
Toward the compact infrared galaxies Arp~220 and NGC~4418, absorption from H$_3$O$^+$ lines tracing the metastable levels of this symmetric top molecule are detected up to energies of 
$\sim 1300$~K. As for the Galactic sources SgrB2 and W31C, these lines
trace the formation process of H$_3$O$^+$ and can therefore be used to accurately infer
the ionization rate which reaches  $\zeta_{\rm H} > 10^{-13}$~s$^{-1}$  in both NGC~4418 and Arp~220,
 as compared with the mean value in the Galactic disk of $\zeta_{\rm H} = 2 \times 10^{-16}$~s$^{-1}$ \citep{indriolo:15}.
 Non metastable lines of H$_3$O$^+$ are
accessible from the ground, the strongest being the $3_2^+ - 2_2^+$ line at 364~GHz  \citep{aalto:11}.

\begin{marginnote}[
]
\entry{AGN}{An Active Galactic Nucleus is a compact region near the center of a galaxy,
 which is emitting intense radiation, as the result of accretion of matter on a super-massive black hole. }
\end{marginnote}

\begin{marginnote}[
]
\entry{Starburst galaxy}{A starburst galaxy is a galaxy forming stars at a rate much higher 
than the average. Instead  of a few M$_\odot$yr$^{-1}$, the star formation rate can reach up to
 10$^3$ M$_\odot$yr$^{-1}$. }
\end{marginnote}

In external galaxies hydrides are important probes of the diffuse clouds, not radiating in CO. 
For instance \cite{appleton:13} report the detection of water vapor in the large shock of the 
Stephan's Quintet, a group of five interacting galaxies. The ground state $p$-H$_2$O transition 
is the brightest submillimeter emission line detected in this system, excluding the FIR structure lines [C{\sc ii}] at 158~$\mu$m and [O{\sc i}] at 63~$\mu$m. The current generation of high resolution spectrometers operating in the visible allows measurement of CH, CH$^+$ and CN
absorption lines towards stars in nearby galaxies. 
In the Magellanic clouds, the relationships between CN, CH, CH$^+$, NaI and KaI are generally consistent with those of the Milky Way but the
overall CH abundance relative to H$_2$ is more variable, likely because of
 the broad range of physical conditions encountered along such sight-lines \rev{and the lower carbon abundance} \citep{welty:06}. 
 The supernova SN2014J in M~82 \citep{ritchey:15} provided a bright enough background
 target to probe the ISM properties in the Galaxy and in M82 along this sight-line.
 The enhanced CH$^+$/CH ratio may indicate an enhanced level of turbulence in this 
 actively star forming galaxy.

\subsection{Active Galactic Nuclei : XDR prototypes}

\begin{marginnote}[
]
\entry{XDR}{X-Ray Dominated Region}
\end{marginnote}

When molecular gas is exposed to intense X-ray radiation, a specific chemistry can develop. As compared with photodissociation regions, X-ray dominated regions (XDR) show longer columns of warm molecular gas because X-rays can penetrate longer columns than UV photons. Because of the increased ionization rate due to 
X-rays, and the higher gas temperature, some molecular abundances are modified.
This is particularly the case for the oxygen hydrides OH$^+$, H$_2$O$^+$ and H$_3$O$^+$. In dense gas illuminated by X-rays, these species are predicted to
present strong emission features, because of the increased excitation by the
high electron density. The close environment of the super-massive accreting black hole in an AGN  can be considered as an X-ray dominated region, even
though some active galaxies also host starbursts. \revv{XDRs also exist in the close 
vicinity of  protostars and have been modeled, for example by  \citet{stauber:05}.} The 
{\it Herschel}/SPIRE spectrum of the Mrk~231 galaxy, an AGN template, 
shows prominent  OH$^+$ and H$_2$O$^+$ rotational lines in emission \citep{vdwerf:10}. 
Other species sensitive to high X-ray irradiation are
atomic carbon, H$_3^+$, H$_2$O and HCN. Because S$^{2+}$ reacts rapidly 
with H$_2$, producing SH$^+$ among other species,  
\cite{abel:08} have proposed that SH$^+$  could be an excellent XDR tracer. This work
remains qualitative because the reaction branching ratio towards SH$^+$ is not known.  A further 
exploration of the reactivity of doubly ionized ions with H$_2$ (e.g. C$^{2+}$)
and of the branching ratios towards hydrides is needed to support this new chemical route.

\subsection{High redshift galaxies}
\begin{figure*}
\includegraphics[width=12cm]{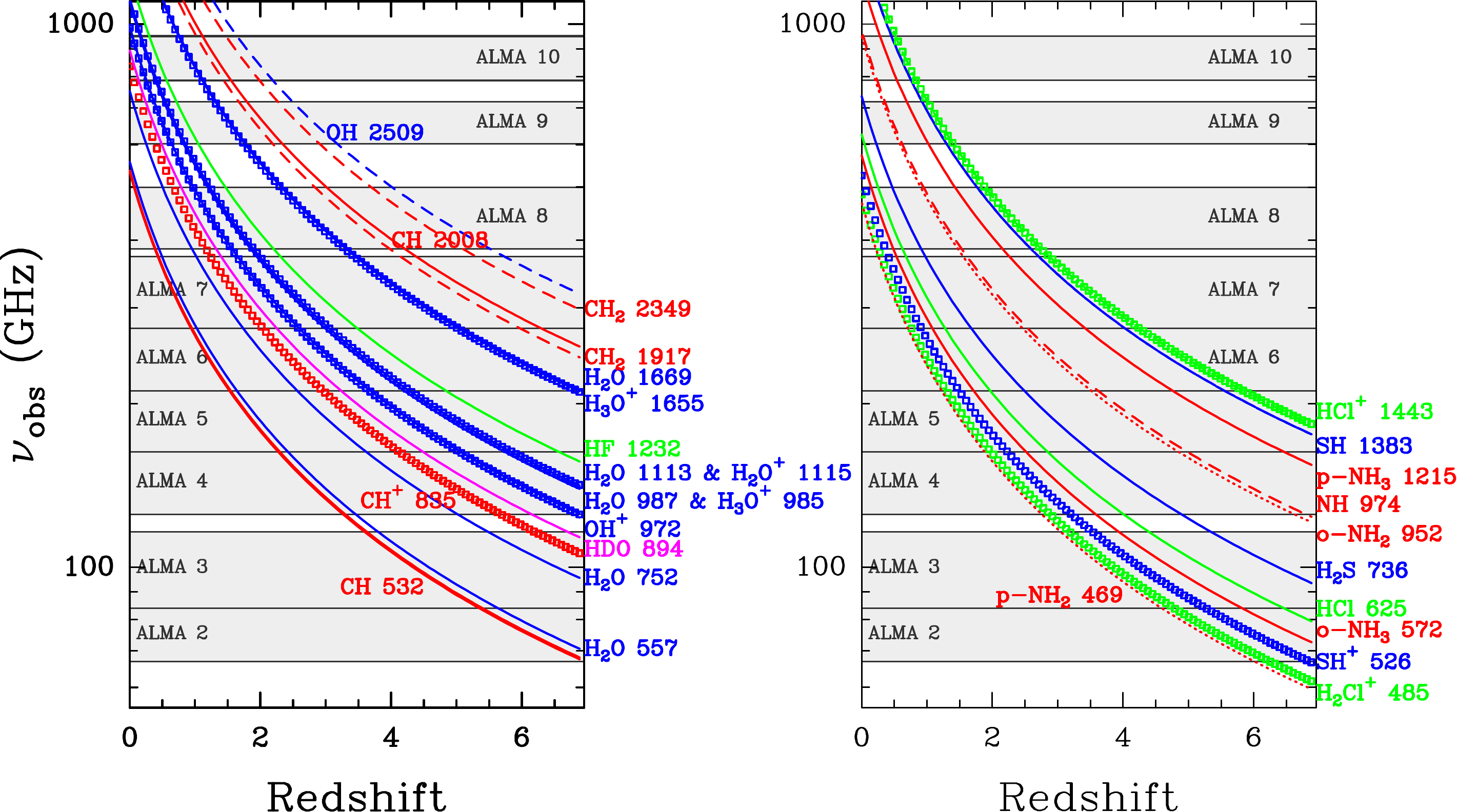}
\caption{Frequencies of the ground state transitions of the main hydrides 
as a function of redshift. Some rotationally excited lines of water vapor are also  included.
 Molecular ions are shown with squares. On the left panel, oxygen-bearing species are displayed in blue,
 carbon-bearing species in red, and halogens in green. On the right panel, nitrogen-bearing species are shown in
 red, sulfur-bearing species in blue and halogens in green.  ALMA bands 2 to 10 are indicated in grey.}
\label{fig:alma}
\end{figure*}

Hydride lines at THz 
rest frequencies are redshifted into the (sub)millimeter range and become accessible to 
the ALMA and NOEMA (NOrthen Extended Millimeter Array) interferometers 
, offering new diagnostic capabilities of the conditions
 in high redshift galaxies. \textbf{Figure \ref{fig:alma}} shows the ground state hydride
 lines accessible to ALMA with the corresponding redshift range. For water vapor, excited
lines are also easily detected \citep{omont:13}, with line strength only a factor two to three 
lower than CO. Nevertheless, the detection of 
water vapor at high redshift was achieved about two decades after that of
 CO, largely because the limited band-width of available receivers prohibited 
searching over a broad frequency ranges. As the CO rotational line frequencies
 are relatively distant from the strongest water line frequencies, 
targeting CO meant avoiding H$_2$O until recently. The tight correlation
of the H$_2$O line luminosity with the FIR luminosity may be used
 to derive the star formation rate.

Towards gravitationally lensed QSOs, the gas of the lensing galaxy may 
produce absorption lines on the continuum of the background object. 
The sight-line towards the radio loud QSO PKS1830-211 is particularly rich, 
with numerous absorption lines detected toward the two main images, sampling 
two lines of sight in the lensing galaxy at z = 0.89. ALMA observations 
enabled the detection of the same hydride lines that were seen in the Milky 
Way by {\sl Herschel} \citep{muller:14a,muller:14b}. Other high redshift hydride detections
 include \rev{H$_2$O$^+$ in a bright submillimeter galaxy \citep{weiss:13}, CH in the average spectrum of 22 such objects 
\citep{spilker:14}}, CH$^+$ in the optical afterglow spectrum of the gamma ray burst GRB~145050A at z=0.89
 \citep{fynbo:14} and in three hyper luminous galaxies \citep{falgarone:15}.
 
  While the same species are detected at $z \sim 1$ as in the Milky Way,  
some hydrides
may become undetectable at higher redshift because of the lack of the 
corresponding
chemical element. For instance HF has been tentatively detected in Cloverleaf QSO at  $z=2.56$ \citep{monje:11} but not in the higher redshift source APM~08279+5255 at $z=3.89$ \citep{lis:11}.
 The absence of HF is best  explained by a low fluorine abundance, since the same source exhibits 
 bright  CO and H$_2$O lines. 

 The comparison of carefully selected hydride line profiles provides an interesting test of 
 the variation of two fundamental constants
with time, the fine structure constant $\alpha = e^2/\hbar c$ and the proto-to-electron mass 
ratio $\mu = m_p/m_e$.  The sensitivity of the CH doublet at 532/536~GHz  to
variations in $\alpha$ and $\mu$ have been computed by \cite{denijs:12}. These  
lines can be observed simultaneously with the  $1_{10}-1_{01}$ transition of $o$-H$_2$O,  
reducing the systematic effects when comparing lines at  very different frequencies, 
and therefore are complementary to other diagnostics like NH$_3$, CH$_3$OH and H$_2$
  discussed by \cite{jansen:14}.

\section{CONCLUSIONS AND FUTURE PROSPECTS}
\label{sec:conclusion}

Since their first discovery in the diffuse interstellar medium, hydrides have been proven 
to exist in the wide variety of interstellar and circumstellar
environments, from the most diffuse, almost purely atomic regions, up to
 dense cores and protoplanetary disks. Given the relative simplicity of
their chemistry, and the good level of understanding of their formation
processes, the diagnostic properties of hydrides need to be further explored
in the future, taking advantage of the increasing capabilities of 
high spectral resolution spectrometers from UV-visible up to centimeter wavelengths.
As shown in Figure \ref{fig:alma}, hydrides ground state
transitions, which are mostly not accessible from  ground based observatories, are redshifted to 
spectral regions with  better atmospheric transmission in distant galaxies, where ALMA and NOEMA 
are operating. This
offers diagnostics of the presence of molecular hydrogen, cosmic ray 
ionization rate,
elemental abundances as well as dynamical processes in the forms of shocks,
turbulence and large scale winds, from the local interstellar medium up to the
most distant galaxies. The combined increase in sensitivity and spectral
resolution will enable more accurate measurements of isotopic ratios, with a
focus on deuterium and nitrogen, as a means to understand the origin of
volatiles in the solar system, and more generally the formation of planets in
circumstellar disks.  It will be exciting to connect the ISM information with 
the measurements performed on the
 67P/Churyumov-Gerasimenko comet by the Rosetta spacecraft.

At longer wavelengths, \rev{the development of  powerful instruments at centimeter wavelengths} 
offers interesting possibilities for accessing the $\Lambda$-doubling transitions of OH (both in
the ground state and in excited levels), CH and the inversion transitions of
NH$_3$, and for comparing the distribution of these species with that of atomic
hydrogen at unprecedented sensitivity and spatial resolution.
At shorter wavelengths, the launch of the JWST,
and the construction of the next generation of ground-based telescopes,
 will provide new tools for  studying  the relation between ices and
  gas phase species, especially water, methane and associated radicals. 

To obtain the best results from these upcoming facilities, several
 scientific issues  remain to be addressed  in the fields of astrochemistry
and molecular physics.
On the astrophysics side, a significant issue is the understanding of non
equilibrium processes on the abundance and excitation of the main
hydrides. As described  above, for a reactive hydride like CH$^+$, 
both the large scale and small scale dynamics of the media, their
multi-phase structure, as well as the details of the formation and excitation
process  all contribute to the production and excitation. A related topic is
the understanding of the hydride OPR, and how these relate to the H$_2$ OPR. This requires investigations of the reaction mechanisms leading to
the hydride formation and destruction, taking the spin symmetry states into
account,  as well as more accurate astronomical measurements.

The field of laboratory astrophysics is developing fast. The interplay between
the gas phase and the solid phase is a key area for progress. With their formation
through hydrogen abstraction reactions, hydride abundances
 are particularly sensitive to rates of the dissociative recombination
of molecular ions. The new cryogenic storage ring (CSR) in Heidelberg should
provide measurements of the dissociative recombination rate
of rotationally cold ions such as OH$^+$. 
Other exciting developments in laboratory astrophysics include 
deeper studies of the interaction of ices with FUV and energetic radiation
using synchrotron radiation sources,  particle bombardment  simulating 
cosmic ray irradiation, and advanced theoretical calculations of
molecular processes. Close collaborations between molecular 
physicists and astronomers will remain a key asset in this area, as it has
been in the past decades.


\section*{ACKNOWLEDGMENTS}
We thank M. Ag\'undez, J. Black, B. Godard, D. Hollenbach, N. Indriolo, E. Falgarone, S. Federman, H. Liszt, P. Sonnentrucker, E. Roueff, and M. Wolfire for a critical reading of an early version of the manuscript, \rev{and our editor E. van Dishoeck for her detailed comments and relevant suggestions.} We owe a special thank to  M. Ag\'undez, E. Dartois, B. Godard, F. Levrier, Z. Nagy, and C. Persson for help with the preparation of some figures.  
The work of D.A.N. has been supported by a grant issued by NASA's Jet
Propulsion Laboratory for the analysis of Herschel data, and a grant from
NASA's Astrophysics Data Analysis Program (ADAP). JRG thanks the Spanish MINECO and the ERC for support under grants AYA2012-32032  and ERC-2013-Syg-610256-NANOCOSMOS. MG thanks the INSU CNRS program PCMI, and CNES for support. Figures \ref{fig:NW09}, \ref{fig_pdr_chp}~left, \ref{fig_pdr_mod_hol}, 
\ref{fig:pacs}, \ref{fig:T08f2} and \ref{fig:ohplus_hist}  are reproduced with permission from AAS managed 
by IOP publishing \copyright AAS. Figures  \ref{fig_pdr_chp}~right, \ref{fig:G14f3a} and \ref{fig:ices}  are reproduced with permission from Astronomy \& Astrophysics, \copyright ESO.


\bibliography{full_15nov} 

\bibliographystyle{ar-style2.bst}


\vspace{1cm}
{\itshape Related resources:} On line services
\begin{enumerate}
\item The astrochymist (http://www.astrochymist.org) provides the list of detected molecules in the interstellar medium with the associated bibliographic references and related resources in astrochemistry.
\item BASECOL,  a database devoted to collisional ro-vibrational excitation of molecules by colliders such as atom, ion, molecule or electron. http://basecol.obspm.fr/
\item Leiden Atomic and Molecular Database (LAMDA), a data base providing the basic atomic and molecular data for molecular excitation calculations. http://home.strw.leidenuniv.nl/~moldata/ 
\item Virtual Atomic and Molecular Data Centre (VAMDC). A portal for 28 databases providing atomic and molecular data. 
http://portal.vamdc.org/vamdc\_portal/home.seam 
\item The Cologne Database for Molecular Spectroscopy (CDMS). A molecular line catalog providing line frequencies, intensities and molecule partition functions. https://www.astro.uni-koeln.de/cdms
\item  JPL molecular spectroscopy. A molecular line catalog providing line frequencies, intensities and molecule partition functions http://spec.jpl.nasa.gov/
\item Splatalogue, a portal to molecular spectroscopy databases. http://www.cv.nrao.edu/php/splat/
\item KInetic Database for Astrochemistry (KIDA). A data base of kinetic data. KIDA provides recommendations for a set of important chemical reactions. http://kida.obs.u-bordeaux1.fr/
\item The UMIST Database for Astrochemistry. a data base of kinetic data. UDFA http://udfa.ajmarkwick.net/
\item The ISM platform provides access to numerical codes and data bases of model results, notably the Meudon PDR code and Paris-Durham shock model. http://ism.obspm.fr
\end{enumerate}

\end{document}